\documentclass[12pt, preprint]{aastex}{}
\def\ie{{\it i.e.\/}}
\def\rf{Eqn.~\ref}
\def\rfs{Eqns.~\ref}
\begin{document}
\newcommand{\pd}[2]{\frac{\partial {#1}}{\partial {#2}}}
\title{Evolution of perturbed accelerating relativistic shock waves}
\author{Giuseppe Palma\altaffilmark{1}, Andrea Mignone\altaffilmark{2,3}, Mario Vietri\altaffilmark{1} and Luca Del Zanna\altaffilmark{4}}
\altaffiltext{1}{Scuola Normale Superiore, Piazza dei Cavalieri 7, 56100 Pisa, Italy}
\altaffiltext{2}{Dipartimento di Fisica Generale ``Amedeo Avogadro'' Universit\`a degli Studi di Torino, Via Pietro Giuria 1, 10125 Torino, Italy}
\altaffiltext{3}{INAF/Osservatorio Astronomico di Torino, Strada Osservatorio 20, 10025 Pino Torinese, Italy}
\altaffiltext{4}{Dipartimento di Astronomia e Scienza dello Spazio, Universit\`a di Firenze, Largo Enrico Fermi 2, 50125 Firenze, Italy}
\begin{abstract}
We study the evolution of an accelerating hyperrelativistic shock under the presence of upstream inhomogeneities wrinkling the discontinuity surface. The investigation is conducted by means of numerical simulations using the PLUTO code for astrophysical fluid dynamics. The reliability and robustness of the code are demonstrated against well known results coming from the linear perturbation theory. We then follow the nonlinear evolution of two classes of perturbing upstream atmospheres and conclude that no lasting wrinkle can be preserved indefinitely by the flow. Finally we derive analytically a description of the geometrical effects of a turbulent upstream ambient on the discontinuity surface.
\end{abstract}
\section{Introduction}{}
There seems to be strong evidences that Gamma-Ray Bursts (GRBs -- see Piran (2005) for a review) involve flows of dense shells thrown by a dying compact star in the ambient medium at Lorentz factors $\Gamma>10^2-10^3$. When the ejecta impact on the surrounding matter carried by a pre-existing stellar wind, a shock is formed and begins to propagate into a decreasing atmosphere, circumstance which leads, for sufficiently steep density profiles, to the shock acceleration. The length scale $k_0^{-1}$ on which the stellar atmosphere rarefies maybe reasonably much smaller than the distance from the center of the star, thus justifying the approximation of planar symmetry in studying the shock evolution.

The problem of a blast wave moving into a decreasing atmosphere has been analyzed both in its Newtonian~\citep{gand,sakurai,raizer,grover,hardy} and relativistic~\citep{bmk,best,perna,naka,pan,sari} regimes, and several self-similar solutions have been found for the flow in both power-law and exponentially shaped density profiles. Despite the importance of the issue, very few papers have been spent to study the stability of the system subject to wrinkling perturbations. In Newtonian regime~\citet{chev1} and~\citet{luo} studied the exponential atmosphere; power-law profiles have been considered in~\citet{sari2}. \citet{wang}, while taking into account relativistic effects, were not able to find any self-similarity in the perturbation analysis of a spherical blast wave propagating in a power-law atmosphere.

\citet{palma} performed a linear stability analysis of a highly relativistic planar shock propagating in an exponential atmosphere and retrieved a self-similar solution for the first order problem. 
They obtained that, at least in the small perturbation limit, with respect to what happens in the Newtonian regime, the corrugation wavelength $k^{-1}$ can drop by a factor of $\Gamma$ still giving rise to no sensible restoring effect in the flow, a behaviour reminiscent of the infinite wavelength case,
    even for small ratios $k_0/k$. 
    This allows the instability of the downstream energy density to persist, thus delaying the saturation phase. 
   
Of course, as the instability exits the linear regime, a numerical approach has to be adopted since the arguments that exclude the arising of stabilizing phenomena in the flow may become weaker and not so pertinent.

The plan of the paper is as follows. In \S\ref{par1} we review the analytical properties of the problem extensively discussed in~\citet{perna} and~\citet{palma}: these predictions provide useful benchmarks to which our numerical scheme can be compared. In \S\ref{par2} we describe the code used to integrate the relativistic hydrodynamics (RHD) equations; a specific subsection is reserved to explain how we overcome the relevant technical difficulties. \S\ref{par3} lists all the major tests through which we run the code before declaring it reliable for our purposes. In \S\ref{par4} we tackle the central subject of the paper, thus reporting several results of simulations dealing with nonlinear variations of the perturbations induced into the system in \S\ref{par3}. Lastly, in \S\ref{par5}, we will show that self-similarity is reached fast enough to allow for an analytical expression for the shock speed in a quite arbitrarily shaped atmosphere. By means of such a result we will develop a technique to calculate the shock position without making time expensive simulations and will apply it to describe the shape evolution of a planar shock impacting a turbulent upstream. Conclusions are drawn in the shape of an excursus in \S\ref{concl}.
\section{Self-similar solution}{}\label{par1}
In this section we summarize the predictions, analytically derived in Perna \& Vietri (2002) and in Palma \& Vietri (2006), whose accuracy we will check in the following. In the first one the self-similar solution for an accelerating hyperrelativistic shock propagating in a planar exponential atmosphere is derived. Assuming an ambient density given by \begin{equation}\rho(x)=\rho_0e^{-k_0x}\,,\end{equation} dimensional and covariance arguments impose the self-similar shock speed $V(t)$ (hereafter we will pose $c=1$) to satisfy \begin{equation}\frac{k_0t}{-\alpha}=\frac{1}{2}\log\frac{(1+V)(1-V_0)}{(1-V)(1+V_0)}-\frac{1}{V}+\frac{1}{V_0}\,,\label{trans}\end{equation} where $V_0$ is the shock speed at time $t=0$ and $\alpha$ is a dimensionless (negative) constant to be determined by imposing a smooth passage of the flow through a critical point. As the shock enters the hyperrelativistic regime, \rf{trans} becomes \begin{equation}\Gamma(t)\approx\Gamma_i\exp\frac{k_0(t-t_i)}{-\alpha}\approx\Gamma_i\left(\frac{\rho}{\rho_i}\right)^{1/\alpha}\,.\label{shloc}\end{equation} Here $\Gamma$ is the shock Lorentz factor and the subscript $i$ refers to the initial condition.

In order to determine the value of $\alpha$ and the downstream profiles of the relevant hydrodynamical quantities, the exact adiabatic fluid flow equations as well as Taub's jump conditions across the shock are considered in their highly relativistic limit. Chosen the self-similarity variable \begin{equation}\xi=k_0[x-X(t)]\Gamma^2(t)\label{csi}\end{equation} ($X(t)$ being the shock position), the hydrodynamics equations can be cast into self-similar form by means of the following separations of variables: \begin{equation}\gamma^2(x,t)=g(\xi)\Gamma^2(t)\,,\quad e(x,t)=q_0R(\xi)\Gamma^{2+\alpha}(t)\,,\end{equation} \begin{equation}\quad n(x,t)=z_0N(\xi)\Gamma^{2+\alpha}(t)\,,\end{equation} with $\gamma$, $e$ and $n$ being, respectively, fluid local Lorentz factor, proper energy density and baryon number density (the first and last ones as seen from the upstream frame), $q_0\equiv\rho_0/\Gamma_i^{\alpha}$ and $z_0\equiv n_0/\Gamma_i^{\alpha}$. Taub's jump conditions are satisfied simply by fixing \begin{equation} g(0)=\frac{1}{2}\,,\quad R(0)=2\,,\quad N(0)=2\,.\end{equation} Solving the equations with respect to $g(\xi)$, $R(\xi)$ and $N(\xi)$ one finds that self-similar quantities satisfy the following Cauchy problem: \begin{equation} R'=\frac{2g\left[-2\alpha(4+\alpha)+(2+\alpha)(\alpha-4\xi)g\right]R}{\alpha^2+(\alpha-4\xi)g\left[-4\alpha+(\alpha-4\xi)g\right]}\,,\label{rp}\end{equation}\begin{equation} g'=\frac{g^2\left[4(\alpha-4\xi)g-14\alpha-3\alpha^2\right]}{\alpha^2+(\alpha-4\xi)g\left[-4\alpha+(\alpha-4\xi)g\right]}\,,\label{gp}\end{equation}\begin{equation} N'=N\frac{2g[(2+\alpha)/\alpha]-g'/g}{g(1-4\xi/\alpha)-1}\,.\label{np}\end{equation} Demanding the simultaneous vanishing of the numerators and denominators of \rfs{rp} and \ref{gp} at a critical point (thus specifying the ``second type'' nature of this self-similar problem) it is possible to find \begin{equation}\alpha=-(2+4/\sqrt{3})\,.\end{equation}

Lying on the zeroth order solution hitherto reviewed, \citet{palma} performed a linear stability analysis with respect to a shock wrinkle of wave number $k$. In the $k/(k_0\Gamma)\ll1$ limit, it is shown that causal phenomena transverse to the shock direction of motion cannot carry disturbances too far. This justifies the approximation of infinite wavelength and independent (zeroth order) evolution of each flow column with slightly perturbed constants in the equation for the shock location. Strictly speaking, \rf{shloc} can be integrated to give \begin{equation} X=t-\frac{\alpha}{k_0}\left(\frac{1}{2\Gamma}\right)^2+c_1\,,\quad\Gamma=\Gamma_i\exp\left(-\frac{k_0t}{\alpha}\right)\,.\end{equation} If we perturb $c_1$ we obtain \begin{equation}\delta X\propto\Gamma^0\,,\end{equation}\begin{equation}\delta e\propto R'(\xi)\Gamma^{4+\alpha}(t)\,,\end{equation}\begin{equation}\delta n\propto N'(\xi)\Gamma^{4+\alpha}(t)\,,\end{equation}\begin{equation}\delta\gamma^2\propto g'(\xi)\Gamma^4(t)\,.\end{equation} Alternatively, perturbing $\Gamma_i$: \begin{equation}\delta X=\delta c_1\propto\Gamma^{-2}\,,\end{equation}\begin{equation}\delta e\propto\left(4R(\xi)+4\xi R'(\xi)-\alpha R'(\xi)\right)\Gamma^{2+\alpha}(t)\,,\end{equation}\begin{equation}\delta n\propto\left(4N(\xi)+4\xi N'(\xi)-\alpha N'(\xi)\right)\Gamma^{2+\alpha}(t)\,,\end{equation}\begin{equation}\delta\gamma^2\propto\left(2g(\xi)+2\xi g'(\xi)-\frac{\alpha}{2}g'(\xi)\right)\Gamma^2(t)\,.\end{equation} It is clear that the first mode is the most severe and thus physically relevant for the instability.

Nevertheless they perform the full perturbation analysis which takes explicitly into account the transverse mixing between adjacent columns, thus also obtaining a complete description for the $y$-component of the four velocity: \begin{equation}\delta u_y\propto g_y(\xi)\Gamma^{s-2}(t)\,,\end{equation} with $s$ being the parameter which selects the strong ($s=3$) or the weak ($s=1$) mode and $g_y$ the self-similar profile satisfying \begin{equation} g_y'=\frac{g'g_y}{2g}+\frac{\left[\alpha R'+4gR(s-3)\right]g_y-i\alpha(k/k_0)\sqrt{g}R_1}{2R\left[(\alpha-4\xi)g-\alpha\right]}\,,\label{gy'}\end{equation}\begin{equation} g_y(0)=-\frac{ik}{\sqrt{2}k_0}\label{gy0}\,.\end{equation} Here $g_y$ is purely imaginary since it is $\pi/2$-shifted with respect to the other perturbations.

In the following two sections we will try to numerically recover near all the theoretical results stated above.
\section{Numerical Setup}{}\label{par2}
Numerical simulations are carried out by solving the equations of number density and momentum-energy conservation, \ie \begin{equation}\label{eq:D}\pd{n}{t}+\vec{\nabla}\cdot\left(n\vec{v}\right)=0\,,\end{equation}\begin{equation}\label{eq:m}\pd{\vec{m}}{t}+\vec{\nabla}\cdot\left(\vec{m}\vec{v}+p\right)=0\,,\end{equation}\begin{equation}\label{eq:E}\pd{E}{t}+\vec{\nabla}\cdot\vec{m}=0\,,\end{equation} where $\vec{v}$ is the fluid velocity, $\vec{m}=nm_ph\Gamma\vec{v}$ and $E=nm_ph\Gamma-p$ are, respectively, the momentum and energy density ($m_p$ being the proton mass).

Proper closure of \rfs{eq:D} -- \ref{eq:E} is specified in the form of an equation of state (EoS), relating the specific enthalpy $h=1+\epsilon+p\Gamma/(nm_p)$ with pressure $p$ and (specific) internal energy of the fluid $\epsilon$. For a relativistic perfect fluid, the desired closure is given by the Synge gas~\citep{synge}. For a single-specie fluid given by a mixture of protons and electrons, the equation of state can be approximated by an analytical expression lately presented in~\citet{mig1} and further discussed in~\citet{mig3}:
    \begin{equation}\label{eq:TM}
      h = \frac{5}{2}\Theta + \sqrt{\frac{9}{4}\Theta^2 + 1} \,,
    \end{equation}
    where $\Theta = p/(nm_p)$ is a temperature-like variable. Compared to the ideal gas EoS with constant adiabatic index $\Gamma_g$ for which
    the enthalpy takes the form $h = 1 + \Gamma_g/(\Gamma_g-1)\Theta$, Eq. (\ref{eq:TM}) yields the correct asymptotic limits for very high ($\Theta\to\infty$)
    and low ($\Theta\to 0$) temperatures, reducing to an ideal EoS with $\Gamma_g=4/3$ and $\Gamma_g=5/3$, respectively.
    In \citet{mig1} it is shown that this expression differs by less than $4\%$ from the theoretical prescription given by the Synge gas.
    Since the equation of state is frequently invoked in the process of obtaining the numerical solution, computational efficiency issues largely entitle to the use of an approximated relation.

The conservation laws (\rfs{eq:D} -- \ref{eq:E}) are solved using the relativistic module available in the PLUTO code~\citep{mig2}. PLUTO is a Godunov-type code offering a variety of computational strategies for the numerical solution of hyperbolic conservation laws in one, two or three dimensions. For an extensive
review of such techniques see~\citep{Marti03} and references therein. Being Riemann-solver based, it is particularly fit for the simulation of high-mach number flows, as it is the case here. For the present application, we employ second-order accuracy in time by using characteristic backtracing (see~\cite{col90}) and linear interpolation with $2^{\rm nd}$ order limited slopes. This scheme yields a one-step time integration by providing time-centered fluxes at zone boundaries,
computed by solving a Riemann problem with suitable time-centered left and right states. For the one- and two-dimensional simulations presented below, we adopt the approximate HLLC Riemann solver of~\citet{mig1b}.
\subsection{The Choice of the Reference Frame}{}
Before presenting our numerical results, we discuss how we faced a number of numerical issues.  

Let us begin by considering the fate of an upstream slab one length scale long: due to the highly relativistic shock compression, it will be roughly resized by a factor $\Gamma^2$. In order to justify the hyperrelativistic approximations assumed above, we would be willing to deal with $\Gamma$ at least as big as 10; even higher Lorentz factor are involved with realistic models of GRBs, wherein the compactness problem solution imposes $\Gamma$ to be one or two orders of magnitude higher.

However, such large Lorentz factors demand an increasingly high resolution, if one wishes to properly capture dynamics of the slab
profile one it enter. This requirement becomes even more severe if the evolution of the perturbations has to be followed accurately.
In this case, in order to overcome spurious numerical fluctuations, a resolution of thousands computational zones per length scale is needed.
From these considerations, we conclude that adopting a static uniform grid would result in extremely inefficient calculations. To overcome this limitation, a reasonable alternative is to resort to adaptive mesh refinement (AMR) techniques, thus providing adequate resolution on the regions of interest. Even in this case, however, we still have to face a subtler problem.

It is known that relativistic shock-capturing codes may suffer from excessive dissipation when a region of fluid with exceedingly large inertia interact with a stationary fluid adjacent to it~\citep{mig1}. This is actually the unfavorable situation we are coping with since, in the upstream rest frame (URF from now on), an ultrarelativistic shock advances in a cold, pressure-less static gas. In this reference frame the jumps of the hydrodynamical variables (in particular the energy density) across the front are maximized, leading to an excessive smearing of the shock profile. The deficiency is inherent to any finite difference method attempting to solve the fluid equations on meshes of finite width. Indeed, even a first-order upwind discretization of the scalar advection equation with linear constant velocity $c>0$, \begin{equation}\frac{u_j^{n+1}-u_j^n}{\Delta t}+c\frac{u_j^n-u_{j-1}^n}{\Delta x}=0\,,\end{equation} shows that $u$ satisfies \emph{exactly} another convection-diffusion problem, namely \begin{equation}\pd{u}{t}+c\pd{u}{x}=\frac{c\Delta x}{2}\left(1-\frac{c\Delta t}{\Delta x}\right)\frac{\partial^2u}{\partial x^2}+O(\Delta x^2)+O(\Delta t^2)\,,\end{equation} where the term in brackets on the right hand side must be positive for stability issues. Thus, roughly speaking, the magnitude of the second derivative provides a rough estimate of the diffusion introduced by the numerical algorithm. The situation does not improve with the employment of higher order methods, since the accuracy reverts to first order in proximity of a discontinuity anyway. This conclusion is supported by several numerical experiments (not shown here) showing that the largest dissipation terms, taken to be proportional to the magnitude of the second derivative of the hydrodynamic variables, result in the frame of the upstream fluid whereas are minimized in the shock frame.

In this respect, it is thinkable to study the shock evolution in its initial instantaneously comoving frame (IICF). In fact, in such an inertial reference frame, shock compression results to have a modest factor of 3 (i.e. an upstream slab of unitary length will be resized by a factor 3). Moreover, in the same frame, due to the favorable Lorentz factor composition law, \begin{equation}\Gamma'=\Gamma\Gamma_0(1+\beta\beta_0)\;,\end{equation} the shock will hardly become hyperrelativistic even in the late acceleration stages.
    This allows to follow the long-term evolution of the downstream self-similar lengths as well as the upstream length scales with a comparable number of points.



The price one pays for reducing in such a drastic way the computational cost consists in a quite complex procedure to recover a snapshot of the system as seen by an observer at rest with respect to the upstream. Due to the relativistic non-absoluteness of simultaneity, we had to make a collage with several (ideally infinitely many) pieces of IICF snapshots, each depicting a particular (ideally infinitely narrow) slab (normal to the $x$-axis) of the flow at a particular IICF time.

In particular, referring to IICF quantities by means of primes, we chose as initial condition a planar shock located at $X_0\equiv X'_0\equiv0$ moving rightward in a grid covering a unitary IICF length along the $x$-axis. We assumed a uniform downstream flow connected to immediately pre-shock upstream by usual Taub's jump conditions: such a choice corresponds to a shock which at $t,t'<0$ propagates in a uniform, cold (hence stationary) atmosphere which, at $x=x'=0$, turns exponential. If the simulation lasts $t'_e$ and we are interested in $x'>x'_L$ region (corresponding, in the URF, to a semi-infinite patch which closely follows on the left the hyperrelativistic motion of the shock) we can only inquire into shock evolution up to $t\le\Gamma_0(t'_e+V_0x'_L)$. Let us consider now an array $\hat{X'}\equiv\left\{x'_i\right\}$ containing the distinct $x$-component of the grid. Lorentz transformed array of $\hat{X'}$ in the URF, \begin{equation}\hat{X}=\frac{\hat{X'}}{\Gamma_0}+\beta_0t\;,\end{equation} contains the $x$-components of the leftmost sector of the patch introduced above. In order to derive hydrodynamical quantities $\hat{Q}$ as measured in the URF at point $x_i$ and time $t$ one has to analyze the snapshot taken in the IICF at time $t'=\Gamma_0(t-\beta_0x_i)$: \begin{equation}\hat{Q}(x_i,t)=\hat{F}(\hat{Q'}(x'_i,t'))\;.\label{tr}\end{equation} Here $\hat{F}$ denotes the map functions transforming density, velocity and pressure from the IICF to URF. 
    Since the time-marching algorithm evolves by discrete time steps, we performed a linear interpolation between the two set of quantities obtained by replacing $t'$ in \rf{tr} with, respectively, $t'^{n-1}$ and $t'^n$ such that $t'^{n-1}\le t'(t,x_i)\le t'^n$.

Such a discussion can be easily extended to an AMR structure, having care to perform temporal interpolation between the finest level step times available at each spatial position.
\subsection{Simulation settings}{}
In all the simulations described in \S\ref{par3} and \S\ref{par4} the initial shock Lorentz factor $\Gamma_0$ is set to 50. Similar results have been obtained by studying the evolution of shocks with different highly relativistic initial Lorentz factors. The upstream density of the atmosphere swept up by the shock spans over two orders of magnitude, with a value of $k_0^{-1}\approx11$. The only exception concerning $\Gamma_0$ is the simulation described in 
    \S\ref{exce} and illustrated in Fig.~\ref{trf}, while in \S\ref{sfec} we followed the shock evolution over $\approx7$ length scales.

In all the 1D simulations described in this paper, an effective resolution of $2.56\times10^5$ grid points has been reached by means of 8 refinement levels on a base grid with $10^3$ cells; the domain box was $[0,1]$.

The 2D simulations were performed on static grids with spatial resolutions indicated in Table~\ref{tabi}.
\clearpage
\begin{deluxetable}{cccccc}\tabletypesize{\footnotesize}\tablewidth{.9\textwidth}
\tablecaption{Physical parameters and resolution adopted in the two-dimensional simulations.}\tablehead{\colhead{$k/k_0$}&\colhead{$\varepsilon$}&\colhead{$rk_0$}&\colhead{$[x_b,x_e]\times[y_b,y_e]$}&\colhead{Resolution}&\colhead{Fig.}}
\startdata$4.8$&$0.5$&$-$&$[0,1]\times[-5,5]$&$2\cdot10^3\times5\cdot10^3$&\ref{d2}\\\noalign{\medskip}$24$&$2$&$-$&$[0,1]\times[-1,1]$&$1.2\cdot10^3\times2.4\cdot10^3$&\ref{r1e0BIS_rho_ux}, \ref{r1e0BIS_rho_04-09-14-19}, \ref{r1e0BIS_uy_04-09-14-19}\\\noalign{\medskip}$4.8\cdot10^2$&$0.5$&$-$&$[0,1]\times[-5\cdot10^{-2},5\cdot10^{-2}]$&$4\cdot10^3\times4\cdot10^2$&\ref{r2e1_rho_01-02-03-04-05-06-07-08-09-10-11_a}, \ref{r2e1_rho_01-02-03-04-05-06-07-08-09-10-11_b}, \ref{r2e1_Uy_01-02-03-04-05-06-07-08-09-10-11_a}, \ref{r2e1_Uy_01-02-03-04-05-06-07-08-09-10-11_b}\\\noalign{\medskip}$4.8\cdot10^2$&$2$&$-$&$[0,1]\times[-5\cdot10^{-2},5\cdot10^{-2}]$&$4\cdot10^3\times4\cdot10^2$&\ref{r2e1BIS_rho_01-02-03-04-05-06-07-08-09-10-11_a}, \ref{r2e1BIS_rho_01-02-03-04-05-06-07-08-09-10-11_b}, \ref{r2e1BIS_Ux_01-02-03-04-05-06-07-08-09-10-11_a}, \ref{r2e1BIS_Ux_01-02-03-04-05-06-07-08-09-10-11_b}\\\noalign{\medskip}$4.8\cdot10^3$&$0.5$&$-$&$[0,1]\times[-5\cdot10^{-3},5\cdot10^{-3}]$&$10^4\times4\cdot10^2$&\ref{r2e2_rho_00-01-02-03-04-05}, \ref{r2e2_uy_07}\\\noalign{\medskip}$-$&$3$&$2.4\cdot10^{-1}$&$[0,1.6]\times[0,2.285714]$&$1.92\cdot10^3\times1.38\cdot10^3$&\ref{r5a3A4_rho_04-09-14-19}, \ref{r5a3A4_ux_04-09-14-19}, \ref{r5a3A4_vort_19}\\\noalign{\medskip}\enddata\\\tablecomments{The domain box is defined by the lower and upper coordinates $[x_b,x_e]$ (in the $x$ direction), $[y_b,y_e]$ (in the $y$ direction). Both the domain box and the resolution refer to the base computational grid used in the code.}\label{tabi}\end{deluxetable}

\clearpage
\section{Code Verification}{}\label{par3}
In the following subsections we will try to recover near all the theoretical results discussed in \S 2. Such an ``exercise'' will provide us a powerful tool to test the code and the analysis procedure itself together with a measure of the reliability of those simulations having no clear theoretical counterpart (nonlinear perturbation, transient to self-similarity).
\subsection{Zero-th order solution: hyperrelativistic regime}{}
Let us consider the simulation of a shock propagating till a fixed point in a homogeneous atmosphere. As it enters a region with an exponentially decreasing density profile, the shock exhibits some inertia and its speed setups to the self similar value (Eq. \rf{shloc}) after a little while, see \S 6.

Incidentally, we would like to point out that such a problem is \emph{totally} scalable with respect to the length scale $k_0^{-1}$: having set $c=1$, we are still free to choose the space (and, therefore, time) measurement unit. As a consequence, all the results we will obtain in this paper are completely independent of the specific value assumed by $k_0$. 

As the shock advances into the stratified atmosphere, it appears possible to study both the spatial profile of the downstream 
hydrodynamical quantities and their temporal evolution.

\begin{figure}[!ht]\begin{center}
\epsscale{1}\plotone{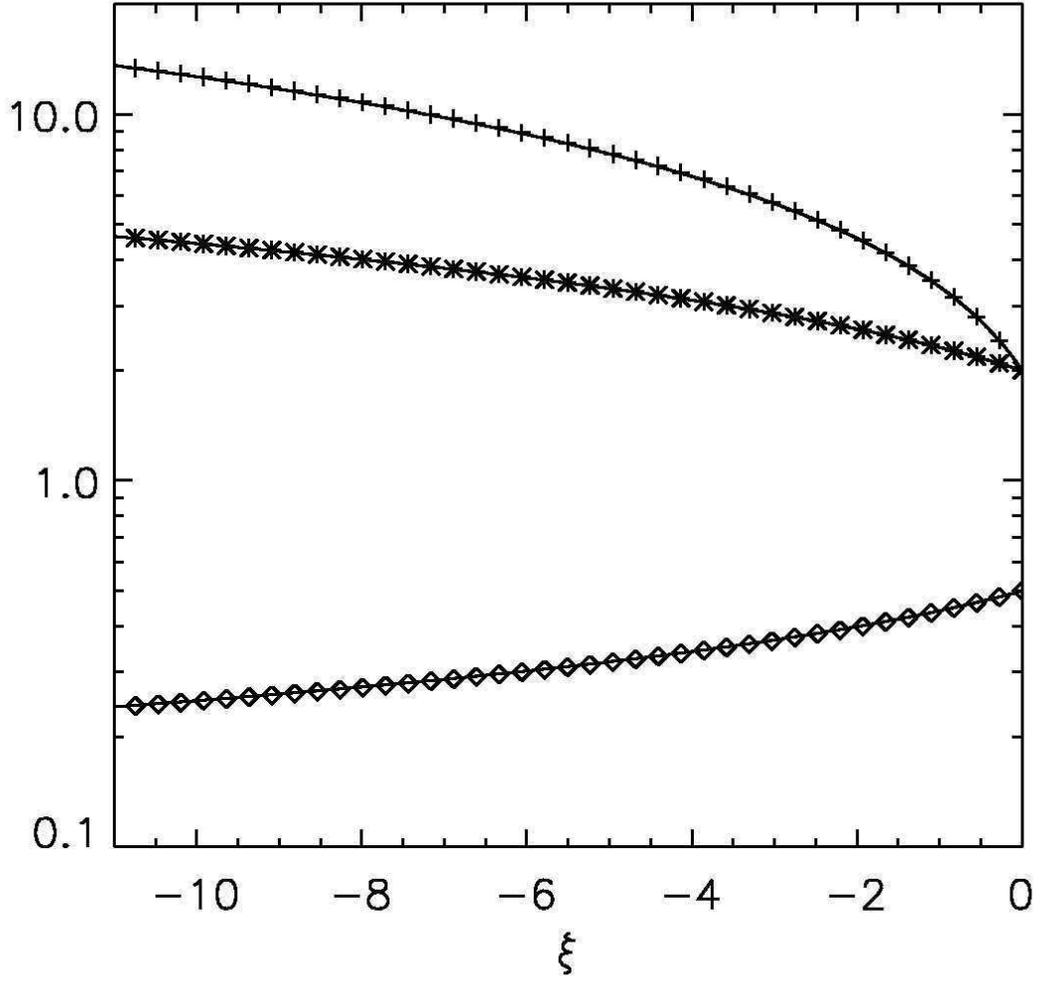}\caption{\footnotesize The figure shows, from top to bottom, the spatial dependence of the dimensionless density $n(\xi)$, pressure $e(\xi)$ and squared Lorentz factor $g(\xi)$: theory predictions (solid lines) are compared with numerical results (40 crosses, stars and diamonds sample in the figure the numerical data).}\label{de}
\end{center}\end{figure}

The snapshot in Fig. \ref{de} shows the density, pressure and Lorentz factor 
(normalized to immediate post shock values of $2$, $2$ and $1/2$) in a small region 
immediately behind the shock front. 
The size of this region is a factor of $\sim 2$ smaller than the length traversed by the point originally 
marking the atmosphere change from homogeneous to exponential.
We also overplot the curves obtained by direct integration of \rfs{rp} -- \ref{np}. As clear
from the plot, we obtain an excellent agreement. 
\clearpage
\clearpage

On the other hand, one can check theoretical rules about Lorentz factor growth as function of time (or, equivalently, under the number of length scale swept up by the shock throughout the simulation). This can be done by means of several, sometimes equivalent, ways. Here we report only two of the most direct methods to be implemented (those our experience suggests to be likely the most robust ones).

A first consistency check can be made by comparing the theoretical value of $\Gamma$ predicted by \rf{shloc} with the one 
computed from our numerical simulations. The latter can be recovered by solving Taub's jump condition with respect to the shock Lorentz factor once pre- and post-shock values have been identified. 

As an alternative, one can consider the value $\xi_L$ of the self-similar variable corresponding, in previous fits, to the leftmost point plotted in Fig~\ref{de}. Strictly speaking, $\xi_L$ must be calculated by numerical inversion, for example, of $R(\xi)$ at the point $R_L$ measured as the leftmost theoretical prediction in plot~\ref{de}. Once $\xi_L$ is known, one can recover the shock Lorentz factor connected to the simulation by inverting \rf{csi}: \begin{equation}\Gamma=\sqrt{\frac{\xi_L}{x_L-X}}\;.\end{equation} A comparison with the usual theoretical value completes the test.

Perhaps it is a point worth of remark the fact that the latter sounds a bit more stringent test than the former, since explicitly assigns a fundamental role to the spatial profile of the downstream in determining the shock speed evolution, thus allowing a more complete point of view on the issue.

Both tests provide an excellent agreement with the related predictions: deviations from theoretical rules appears to be nothing but numerical noise and at almost any time $t>0$ remain below a small fraction, typically less than $1\%$.

As a measure of the code reliability we report what simulations predict about the parameter $\alpha$. From \rf{shloc} \begin{equation}\alpha=\frac{\log\frac{\rho}{\rho_i}}{\log\frac{\Gamma}{\Gamma_i}}\;;\end{equation} substituting simulation values and averaging on several snapshot times we obtain \begin{equation}\alpha=-4.3102\ldots\;,\end{equation} a value which differs from the correct one for less than $0.02\%$.

\subsection{Zero-th order solution: transrelativistic regime}
\begin{figure}[!ht]\begin{center}
\epsscale{.8}\plotone{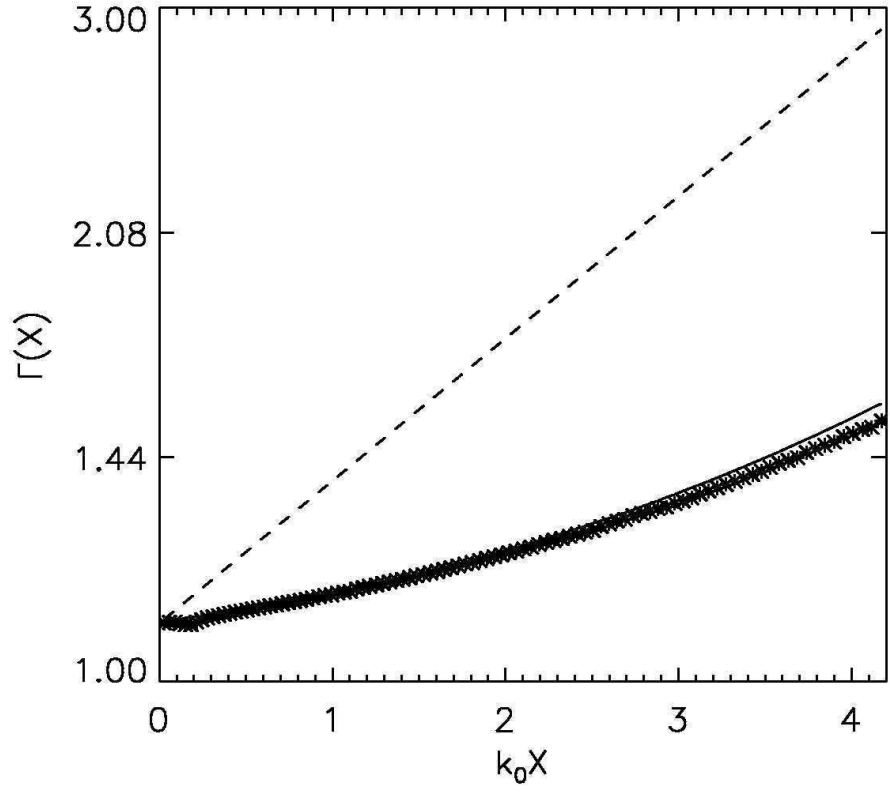}\caption{\footnotesize Evolution of the shock Lorentz factor as obtained by simulation (stars) together with exact self-similar solution (solid line) and its hyperrelativistic approximation (dashed line).}\label{trf}
\end{center}\end{figure}

{}\label{exce}
In this subsection we focus on the only prediction we have as long as the shock is neither Newtonian nor hyperrelativistic: the rule about shock speed given by \rf{trans}. We consider a shock with an initial Lorentz factor $\Gamma_0=1.1$ and follow its evolution for the crossing of $\approx4$ length scales. In Fig.~\ref{trf} we plot $\Gamma$ as a function of traversed length (in units of $k_0^{-1}$), together with the exact self-similar prediction (\rf{trans}) and the run as expected if the shock would have been highly relativistic. 

The excellent agreement we observe in this plot completes the thorough picture about the zero-th order problem.
\clearpage
\clearpage
\subsection{First order solution}{}
In principle, the linear perturbation analysis of a planar shock is a fully 2D problem~\citep{chev1,palma}. However, in order to approach the problem from a numerical point of view, it is convenient to take advantage of the infinite wavelength approximation, whose applicability in the hyperrelativistic regime has been discussed in detail in~\citep{palma}. Such a scheme enables an almost full investigation of the physically relevant phenomena, still allowing a reduced numerical cost.

The idea can be summarized as follows. Firstly we perform the usual 1D simulation of a planar, unperturbed shock wave, as described in previous sections. Then we carry out a second 1D run by perturbing the upstream region with an overdense (by a factor $\varepsilon$) bar, limited in extension to a fraction $\Delta$ of length scale (hereafter $\Delta\approx1.32$); since also this last simulation is 1D, the reader should imagine such a bar indefinitely extended perpendicularly to the shock speed. Moreover -- in order to avoid spurious structures, here as well as in the following section -- we joined smoothly the perturbing bar to the background up to the 4th derivative by means of the factor $\cos^4[\pi k_0(x-\bar{x})/\Delta]$, $\bar{x}$ being the center of the bar (hereafter $\bar{x}\approx0.79k_0^{-1}$), thus obtaining the following upstream density profile:
\begin{equation}\Pi(x)=\rho_0e^{-k_0x}\cdot\left\{1+\varepsilon\cos^4\left[\frac{\pi k_0(x-\bar{x})}{\Delta}\right]H\left(x+\frac{\Delta}{2k_0}-\bar{x}\right)H\left(\bar{x}+\frac{\Delta}{2k_0}-x\right)\right\}\;,\end{equation} where $H$ is the Heaviside step function. Perturbations to hydrodynamical quantities are simply obtained by subtracting term-to-term the values of the second set from the first ones. No 2D simulations were needed and the original resolution along the $x$-axis of the zero-th order analysis has been maintained.

Nonetheless, we warn the reader that any spurious oscillation that might marginally affect the zero-th order profiles, will result here in a severe noise, when trying to recover a variable value as difference between two slightly different quantities. This justifies such a high resolution which prevents us from performing directly a 2D simulation: only this way we can keep the noise down to a reasonable threshold. In order to fix this problem we smoothed the data by performing a local regression using weighted linear least squares with a 2nd degree polynomial. In this way we were allowed to recover the physically relevant development of the perturbation by smoothing away less than 5\% noise.

As in the first subsection, the work can be split up into two stages: the study of the spatial perturbation profile and the test of correct temporal growth.

\begin{figure}[!ht]\begin{center}
\epsscale{1}\plotone{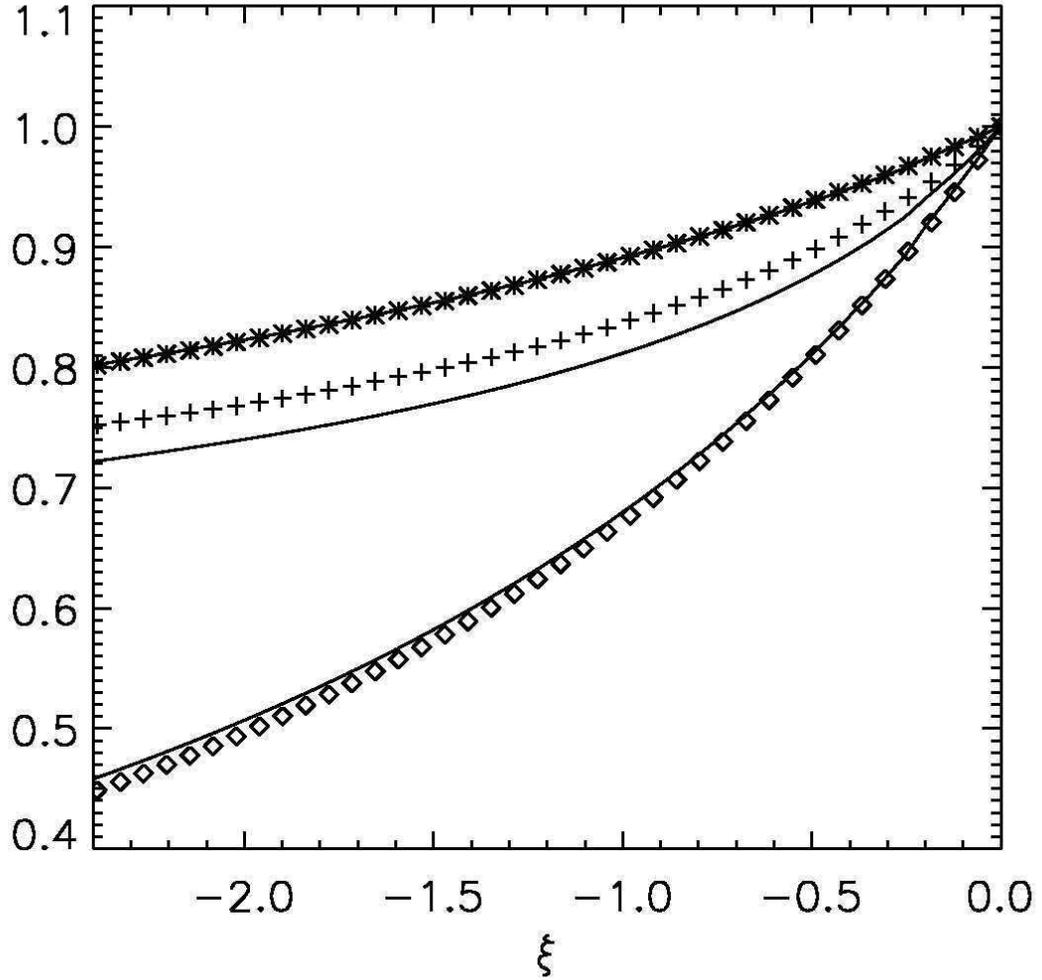}\caption{\footnotesize The figure shows, from top to bottom, the spatial dependence of the perturbations to pressure $\delta e(\xi)$, density $\delta n(\xi)$ and squared Lorentz factor $\delta\gamma^2(\xi)$ normalized to the immediate downstream value: theory predictions (solid lines) are compared with numerical results (40 stars, crosses and diamonds sample in the figure the numerical data).}\label{pd}
\end{center}\end{figure}

Results for the spatial dependence of perturbations of the hydrodynamical quantities (at a fixed time) are reported in Fig.~\ref{pd}, according to theory predictions and to our numerical scheme. The agreement is acceptable, specially considering the way such numerical results are achieved.
\clearpage
\clearpage

\begin{figure}[!ht]\begin{center}
\epsscale{.8}\plotone{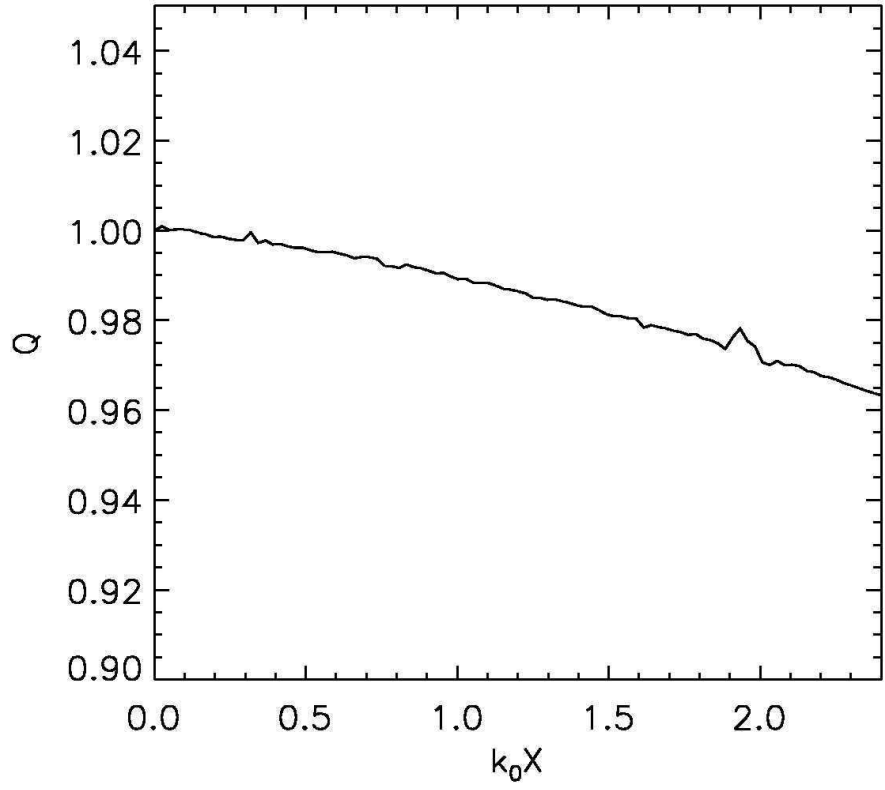}\caption{\footnotesize Temporal evolution of perturbation amplitude normalized to the expected growth. $Q$ is defined as the ratio between simulated and expected perturbation growth; consequently, any gap between theory and numerical test should have as a counterpart a departure of $Q$ from 1, commensurate with the gap.}\label{tg}
\end{center}\end{figure}

Concerning the instability rate, it is possible to compare the temporal dependence of theoretical and numerical perturbations by studying how the simulated amplitude -- normalized to the expected growth -- diverges from unity. In Fig. \ref{tg} the results of such a study are reported: only a minor gap of few percentage points appears, thus stressing once again the good suitability of our scheme even for the subtle task of studying wrinkle perturbations to an hyperrelativistic shock.
\clearpage
\clearpage
\subsection{Finite wavelength wrinkles}{}
Here we just report the results of a 2D simulation (obviously much less resolved in $x$-direction than the 1D previous ones) dealing with an overdense upstream bar not uniformly extended along the $y$-axes as in the previous subsection. Instead, we impose a finite wavelength ($\lambda=2\pi k^{-1}$) sinusoidal profile (along with periodic $y$-boundary conditions), thus allowing in the following section a direct comparison with the effects of an analogous bar inducing nonlinear perturbation to the system.

\begin{figure}[!ht]\begin{center}
\epsscale{.8}\plotone{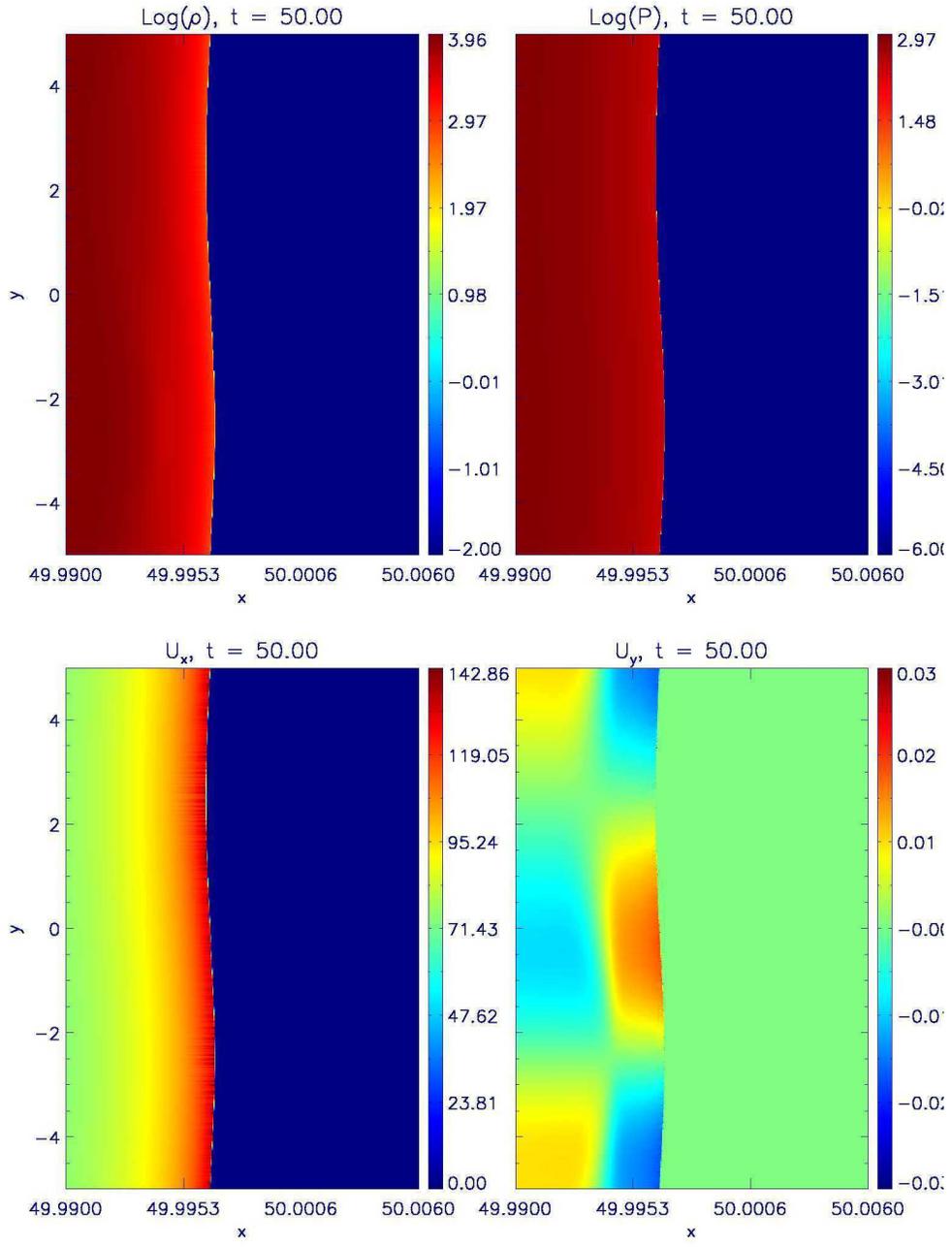}\caption{\footnotesize Hydrodynamical quantities in the linear perturbation regime ($k\approx4.8k_0;\varepsilon\approx0.5;\Delta\approx1$).}\label{d2}
\end{center}\end{figure}

If $k\approx4.8k_0$, the effect of a bar of amplitude $\varepsilon\approx0.5$ and extended $\Delta$ are reported in Fig.~\ref{d2}: we emphasize that the finiteness of the wavelength imply a non-zero $y$ component of the velocity.
\clearpage
\clearpage

In particular it appears noteworthy that the downstream profiles of $u_y$ in Fig.~\ref{d2}, although not sufficiently refined to test the code strictly speaking, are nevertheless completely consistent with theoretical predictions given by \rfs{gy'} -- \ref{gy0} for the strong mode $s=3$.
\section{Nonlinear perturbations}{}\label{par4}
Once the robustness and consistency of the scheme has been demonstrated, we are allowed to study 2D problems heavily involving completely new phenomena or, at least, processes neglected in the small perturbation regime.

In the following we present firstly an idealized problem, aimed at inquiring whether or not the instability reaches any sensible saturation point: the weak sinusoidal bar discussed in the previous section will be replaced by a more substantial one. To follow we will show what happens if a dense cylindrical cloud hampers the downhill path of the shock.
\subsection{Sinusoidal bars}{}
Aiming to study the nonlinear phase of shock perturbations, we will impose here the sinusoidal profile of the perturbing upstream on the density logarithm (\ie~$\Pi(x,y)=\rho(x)\cdot10^{\varepsilon\sin ky}$) rather than on the density itself -- as done previously. In this way we are allowed to use larger amplitude perturbations which often imply a contrast of several orders of magnitude between overdense regions and adjacent vacua. As usual, in all the following simulations, the sinusoidal bars have a width $\Delta k_0^{-1}$.

\clearpage
\begin{figure}[!ht]\begin{center}
\epsscale{.8}\plotone{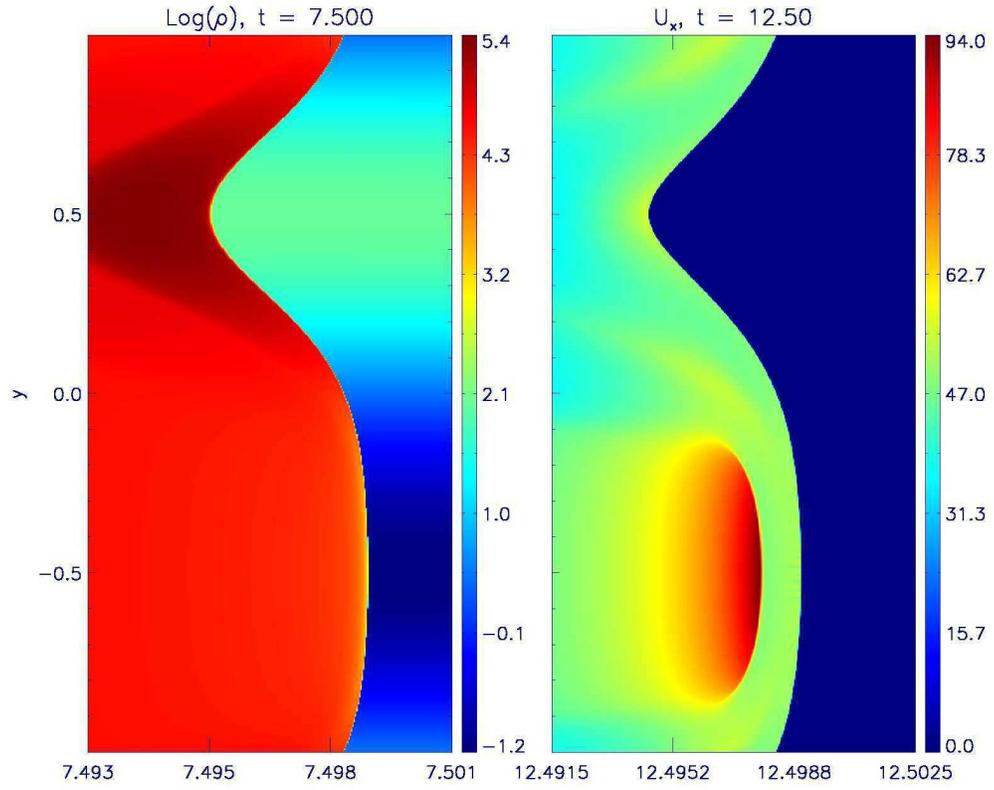}\caption{\footnotesize The density logarithm (left) and the parallel 4-velocity $u_x$ (right) for $k\approx24k_0$ and $\varepsilon\approx2$. Please note that each panel refers to a different time, according to the different phases discussed in the text.}\label{r1e0BIS_rho_ux}
\end{center}\end{figure}
\clearpage

Let us begin by considering a long wavelength density bar ($k\approx24k_0$) with an amplitude of about two orders of magnitude ($\varepsilon\approx2$). Such an upstream inhomogeneity induces a strong corrugation (see left panel in Fig.~\ref{r1e0BIS_rho_ux}, depicting the barionic density in a late phase of the bar shocking process) and highly nonlinear perturbations to downstream flow. In Fig.~\ref{r1e0BIS_rho_ux}, for instance, the right panel shows $u_x$ just after the shock emerges from the bar. The high-speed blob we observe just behind the crest is the relic of the flow corresponding to the acceleration phase in the bar vacuum. Once the shock emerges from the low-density region, the impact on the unperturbed atmosphere produces the reverse shock which separates the fast blob from the main discontinuity.

\begin{figure}[!ht]\begin{center}
\epsscale{.8}\plotone{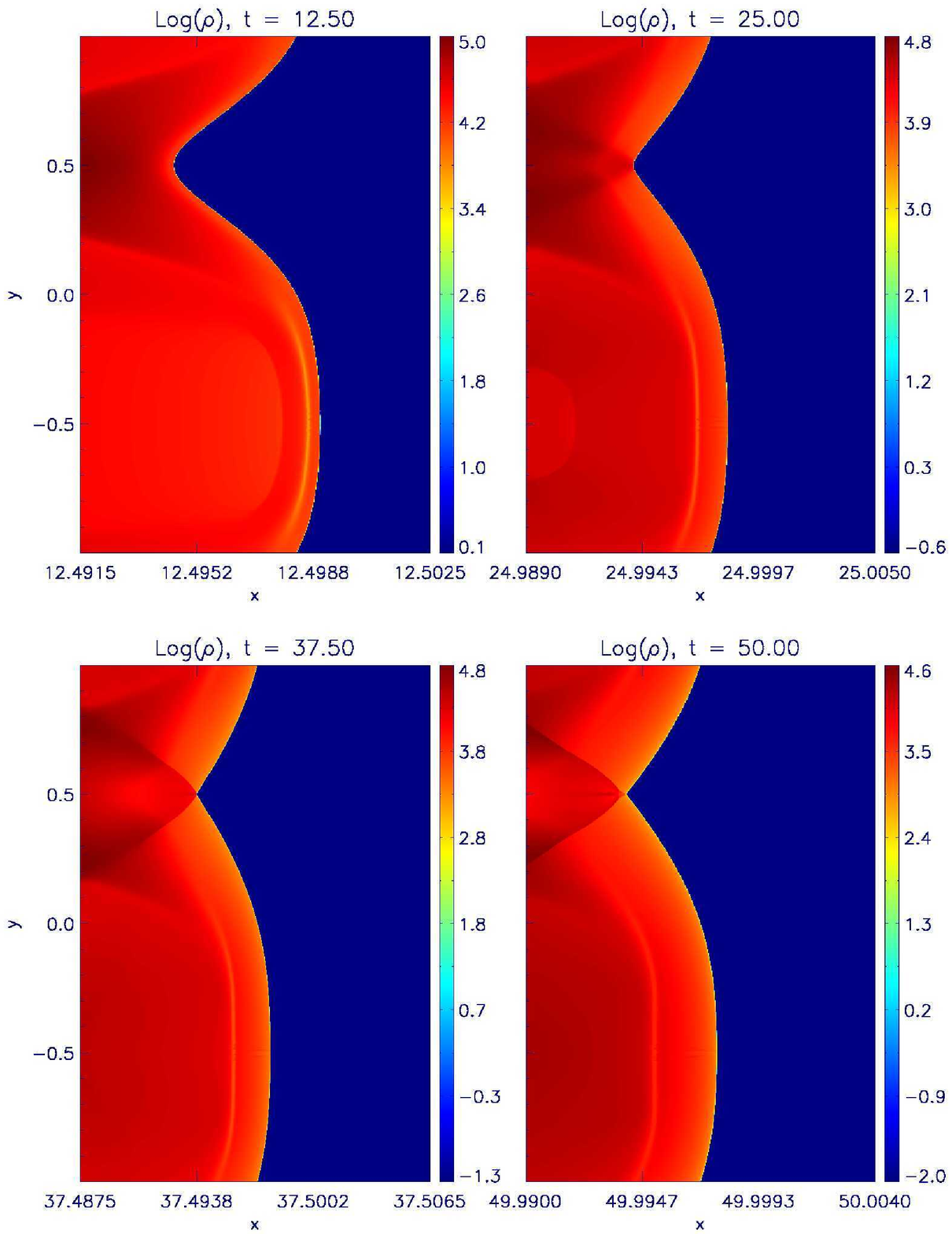}\caption{\footnotesize The density logarithm for $k\approx24k_0$ and $\varepsilon\approx2$.}\label{r1e0BIS_rho_04-09-14-19}
\end{center}\end{figure}

\begin{figure}[!ht]\begin{center}
\epsscale{.8}\plotone{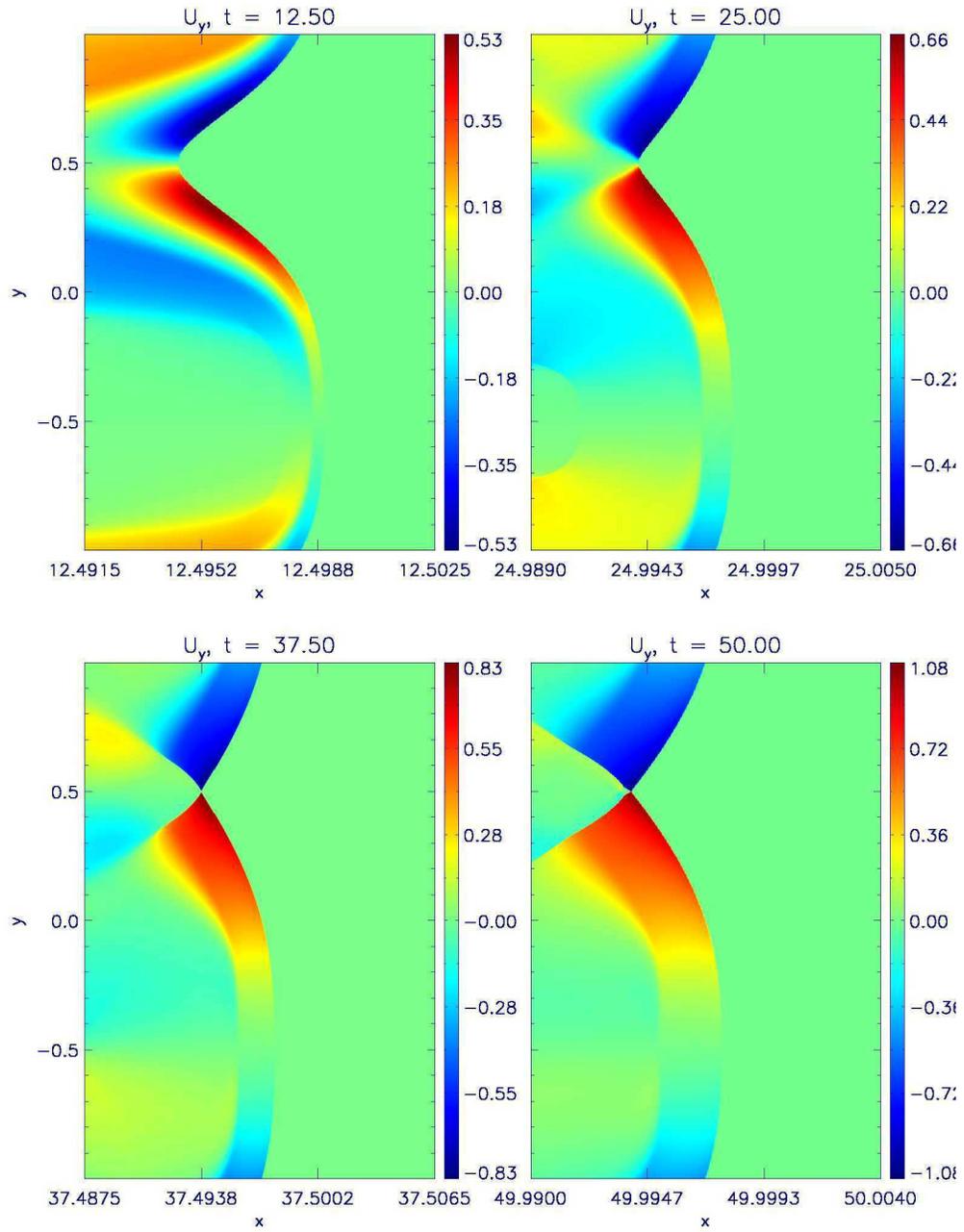}\caption{\footnotesize The transverse 4-velocity $u_y$ for $k\approx24k_0$ and $\varepsilon\approx2$.}\label{r1e0BIS_uy_04-09-14-19}
\end{center}\end{figure}

Moreover, similarly to the linear case discussed in \S 4.4, the shock tends to fill the valleys present in its profile simply by means of something like a potential flow of matter along the discontinuity normal: Fig.~\ref{r1e0BIS_uy_04-09-14-19} shows that the matter flows from the crest to the valley. However, at least on the explored time scales, such a long perturbation wavelength prevents almost perfectly the gap between crest and valley from a quick damping that the mechanism above described would cause to higher wave-number wrinkles (see the sequence of snapshots in Fig.~\ref{r1e0BIS_rho_04-09-14-19}).
\clearpage
\clearpage

At shorter wavelengths, the system evolves in a quite different way.
\clearpage
\begin{figure}[!ht]\begin{center}
\epsscale{.8}\plotone{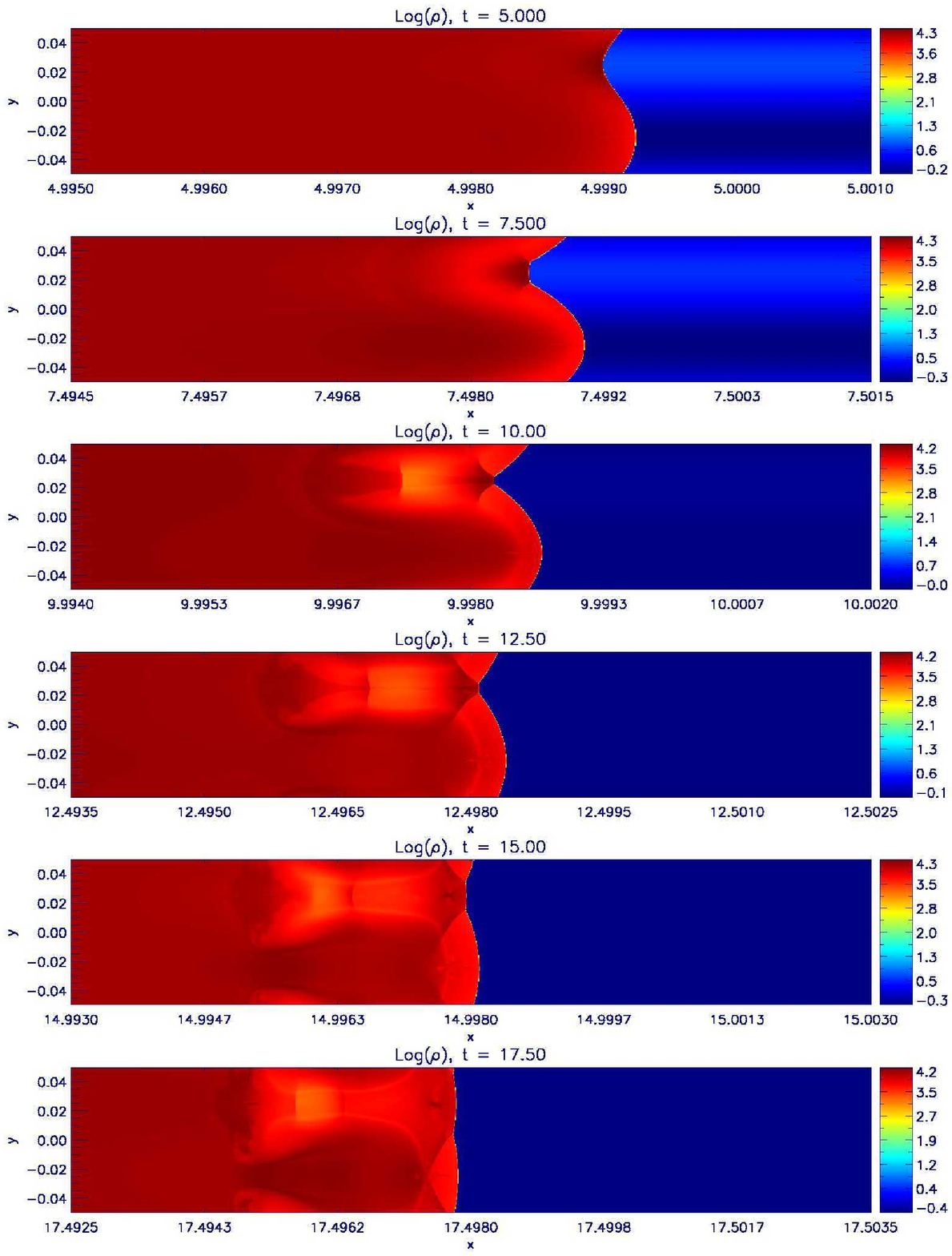}\caption{\footnotesize The density logarithm for $k\approx4.8\cdot10^2k_0$ and $\varepsilon\approx0.5$.}\label{r2e1_rho_01-02-03-04-05-06-07-08-09-10-11_a}
\end{center}\end{figure}
\begin{figure}[!ht]\begin{center}
\epsscale{.8}\plotone{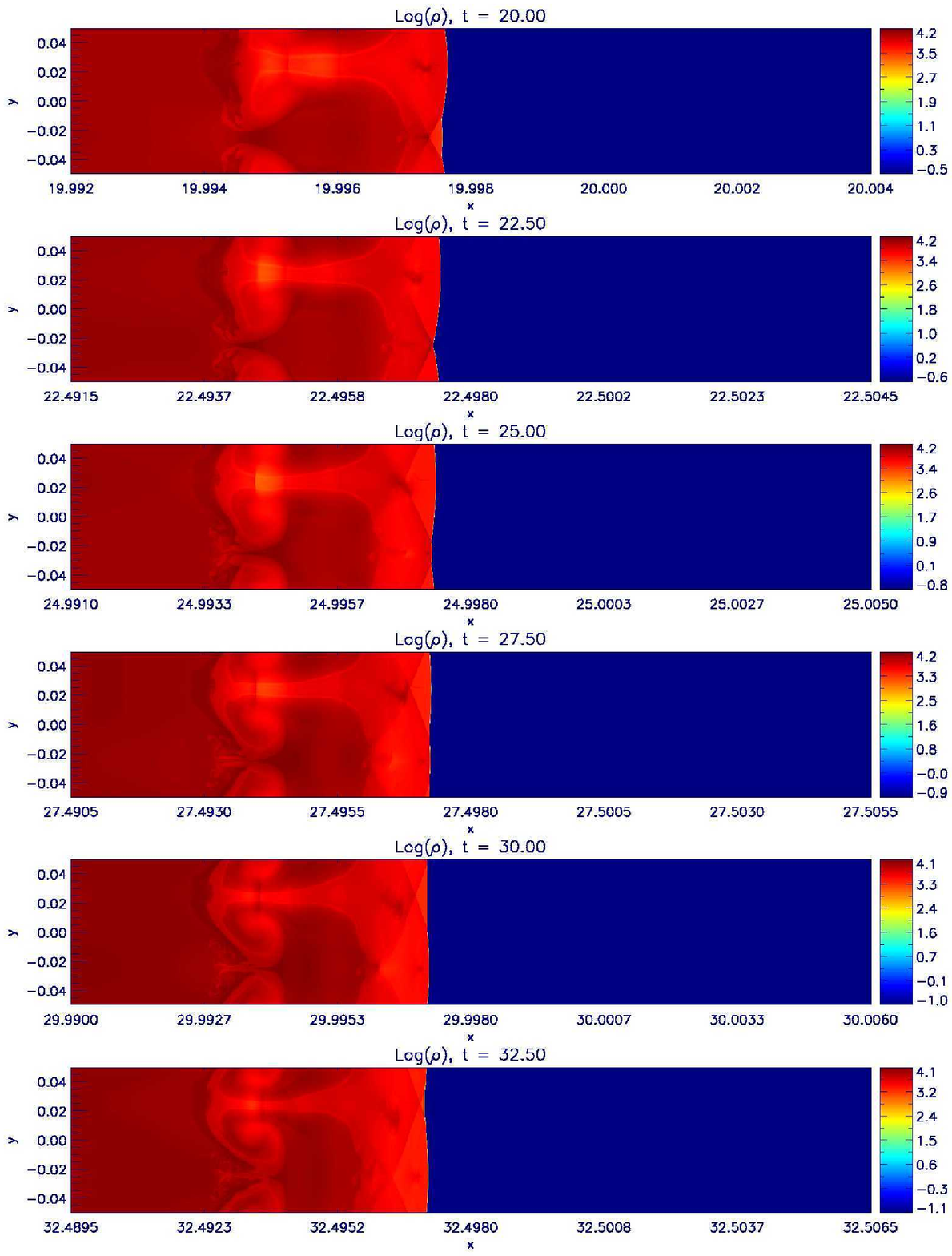}\caption{\footnotesize The density logarithm for $k\approx4.8\cdot10^2k_0$ and $\varepsilon\approx0.5$.}\label{r2e1_rho_01-02-03-04-05-06-07-08-09-10-11_b}
\end{center}\end{figure}
\begin{figure}[!ht]\begin{center}
\epsscale{.8}\plotone{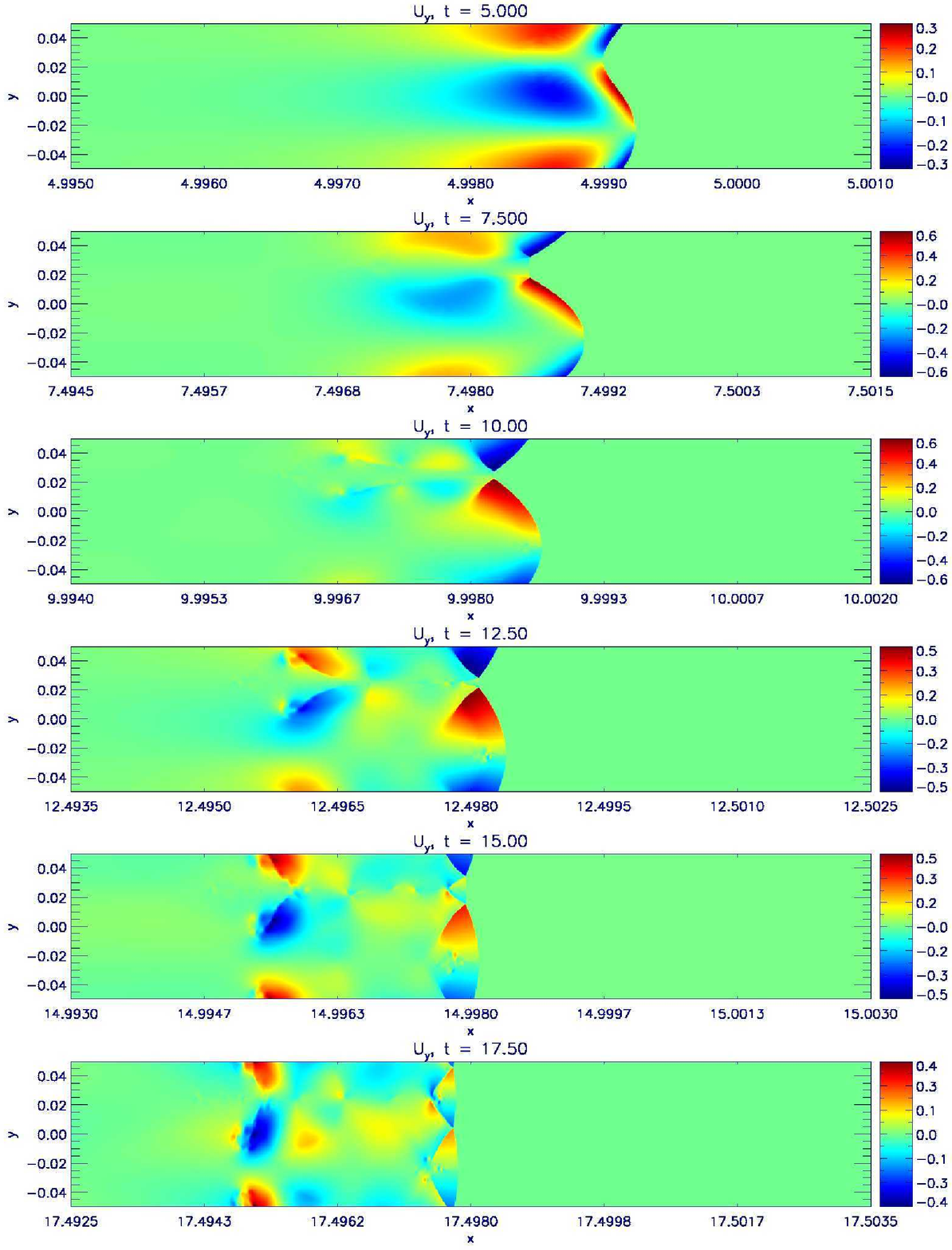}\caption{\footnotesize The transverse 4-velocity $u_y$ for $k\approx4.8\cdot10^2k_0$ and $\varepsilon\approx0.5$.}\label{r2e1_Uy_01-02-03-04-05-06-07-08-09-10-11_a}
\end{center}\end{figure}
\begin{figure}[!ht]\begin{center}
\epsscale{.8}\plotone{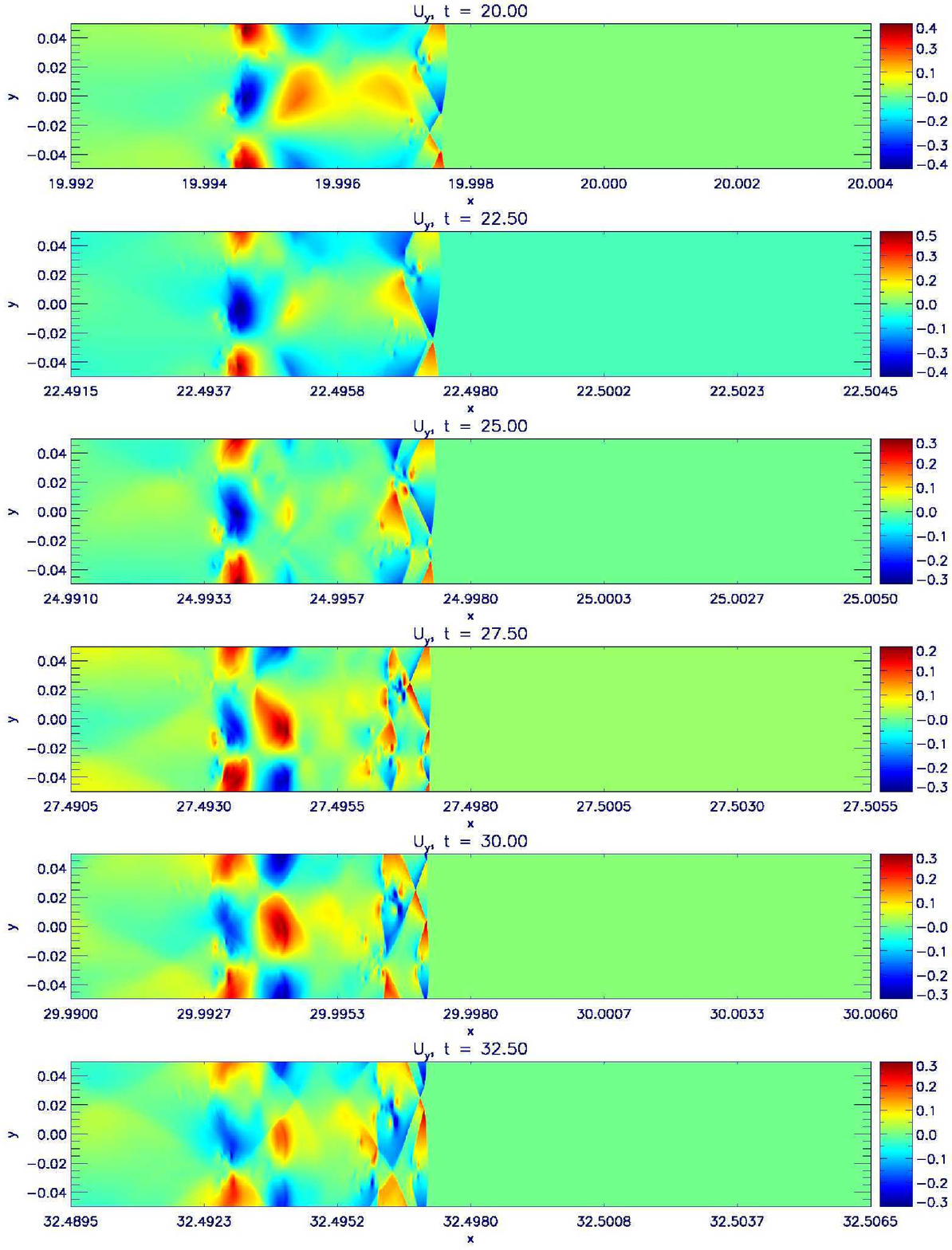}\caption{\footnotesize The transverse 4-velocity $u_y$ for $k\approx4.8\cdot10^2k_0$ and $\varepsilon\approx0.5$.}\label{r2e1_Uy_01-02-03-04-05-06-07-08-09-10-11_b}
\end{center}\end{figure}
\clearpage

Let us focus our attention on a bar of $k\approx4.8\cdot10^2k_0$ and $\varepsilon\approx0.5$. 
One can argue that, due to the reduced transverse distance between equally out of phase flow columns (or, equivalently, thanks to the higher gradients involved), the evolution observed in the previous run resembles the present case in slow-motion: the more $k/k_0$ grows, the faster the evolution gets.
In fact, looking at the sequences of snapshot in Figs.~\ref{r2e1_rho_01-02-03-04-05-06-07-08-09-10-11_a} -- \ref{r2e1_rho_01-02-03-04-05-06-07-08-09-10-11_b} and \ref{r2e1_Uy_01-02-03-04-05-06-07-08-09-10-11_a} -- \ref{r2e1_Uy_01-02-03-04-05-06-07-08-09-10-11_b} which depict the time evolution respectively of barionic density and $u_y$, it is possible to see that a situation similar to the last snapshots in both Figs.~\ref{r1e0BIS_rho_04-09-14-19} and~\ref{r1e0BIS_uy_04-09-14-19} here is reached on a reduced time scale. What we observe in all its progression is the sharpening of the valley, which starts being U-shaped and evolves in the shape of a V. At that point in practice we have two distinct shocks which go to intersect, with the resulting formation of two secondary shocks in the downstream. Such a X-shaped structure evolves with the secondary shocks which advance toward the adjacent crests. In this context, the valley closes the gap on the crest in the short and within another few fractions of length scale comes to overtake the leading front of the shock. At this point the play of the flow columns repeats with reversed roles. However it should be plain that, with each role reverse, two main things happen: first, a new layer is added to the existing pattern of hydrodynamical fluctuations the flow advects far in the downstream; second, the wrinkle amplitude gets smaller and smaller, thus coming to restore the original zero-th order solution. Having these facts in mind, it is possible to give an estimate of the time $T_{sm}$ needed by the shock, once it comes out of the perturbing region, to restore the original planar shape. \citet{palma} most clearly discussed the scaling of such a time, so that we can say $T_{sm}\propto\Gamma/k$. The coefficient of proportionality can be estimated from the simulation: if we say that planarity is restored in the last snapshot in Fig.~\ref{r2e1_rho_01-02-03-04-05-06-07-08-09-10-11_b} and remember that the shock comes out of the perturbing bar (of width $\Delta k_0$) at $t\approx11$, we obtain \begin{equation}T_{sm}\sim25\frac{\Gamma}{k}\;.\label{tsm}\end{equation}

\clearpage
\begin{figure}[!ht]\begin{center}
\epsscale{.8}\plotone{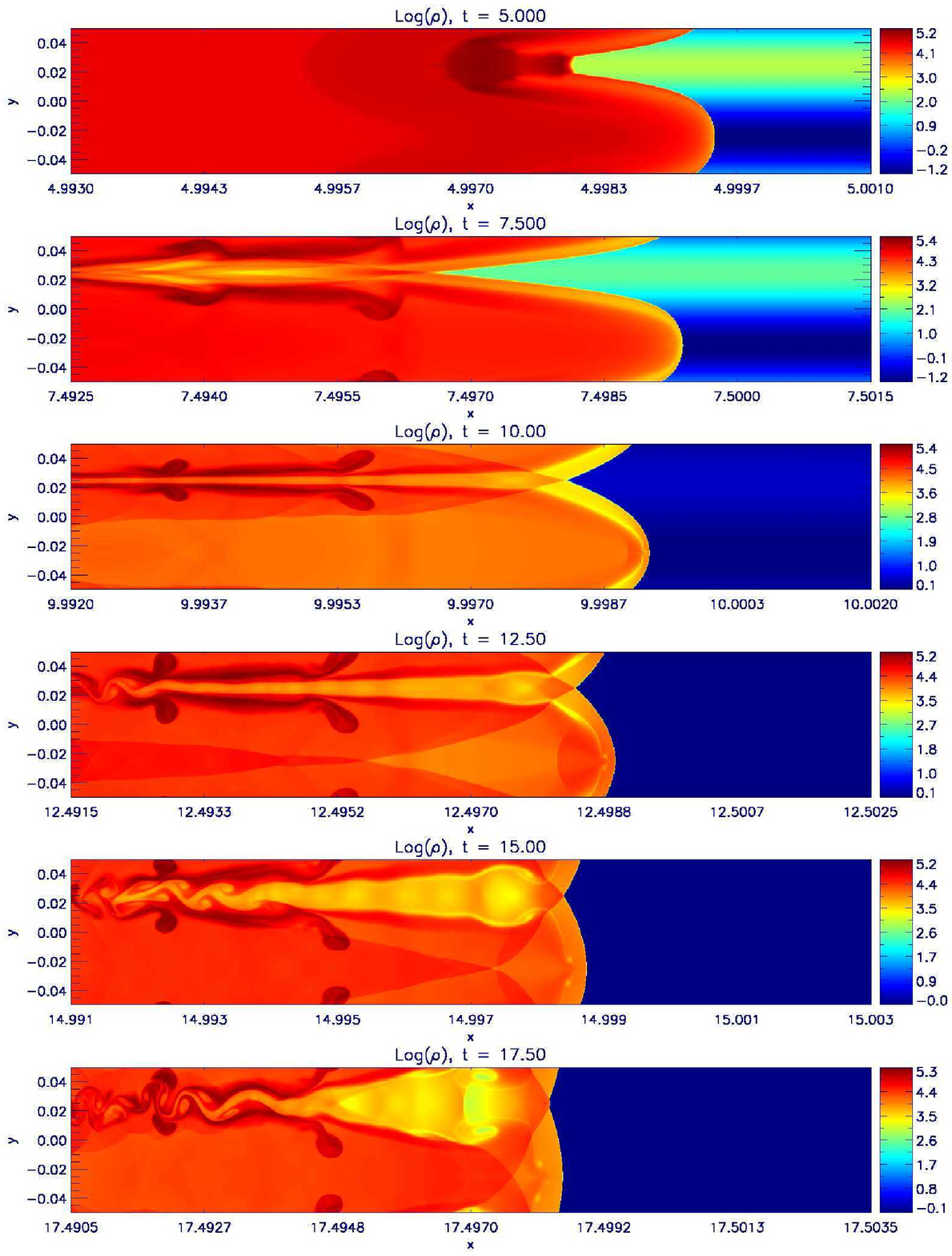}\caption{\footnotesize The density logarithm for $k\approx4.8\cdot10^2k_0$ and $\varepsilon\approx2$.}\label{r2e1BIS_rho_01-02-03-04-05-06-07-08-09-10-11_a}
\end{center}\end{figure}
\begin{figure}[!ht]\begin{center}
\epsscale{.8}\plotone{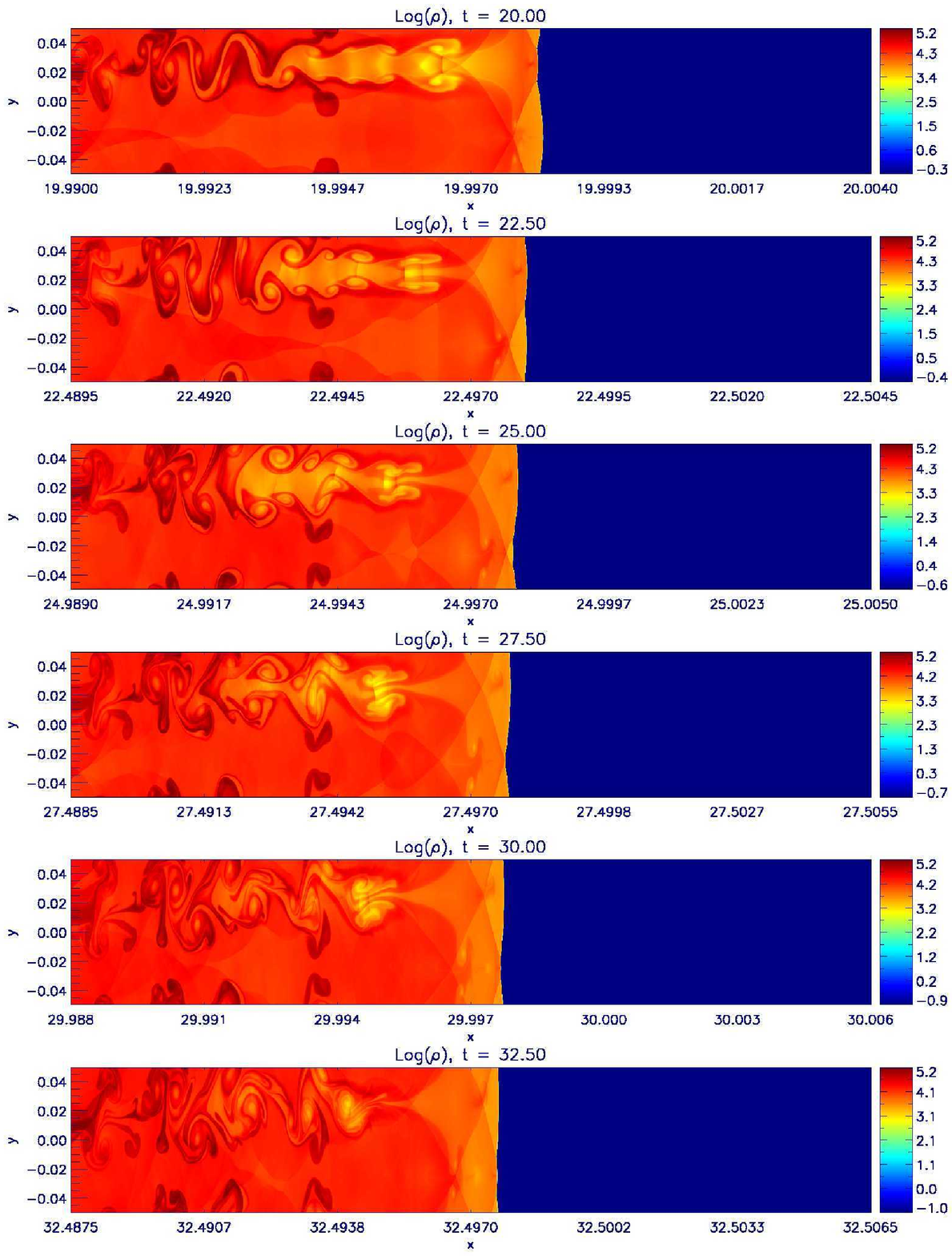}\caption{\footnotesize The density logarithm for $k\approx4.8\cdot10^2k_0$ and $\varepsilon\approx2$.}\label{r2e1BIS_rho_01-02-03-04-05-06-07-08-09-10-11_b}
\end{center}\end{figure}
\begin{figure}[!ht]\begin{center}
\epsscale{.8}\plotone{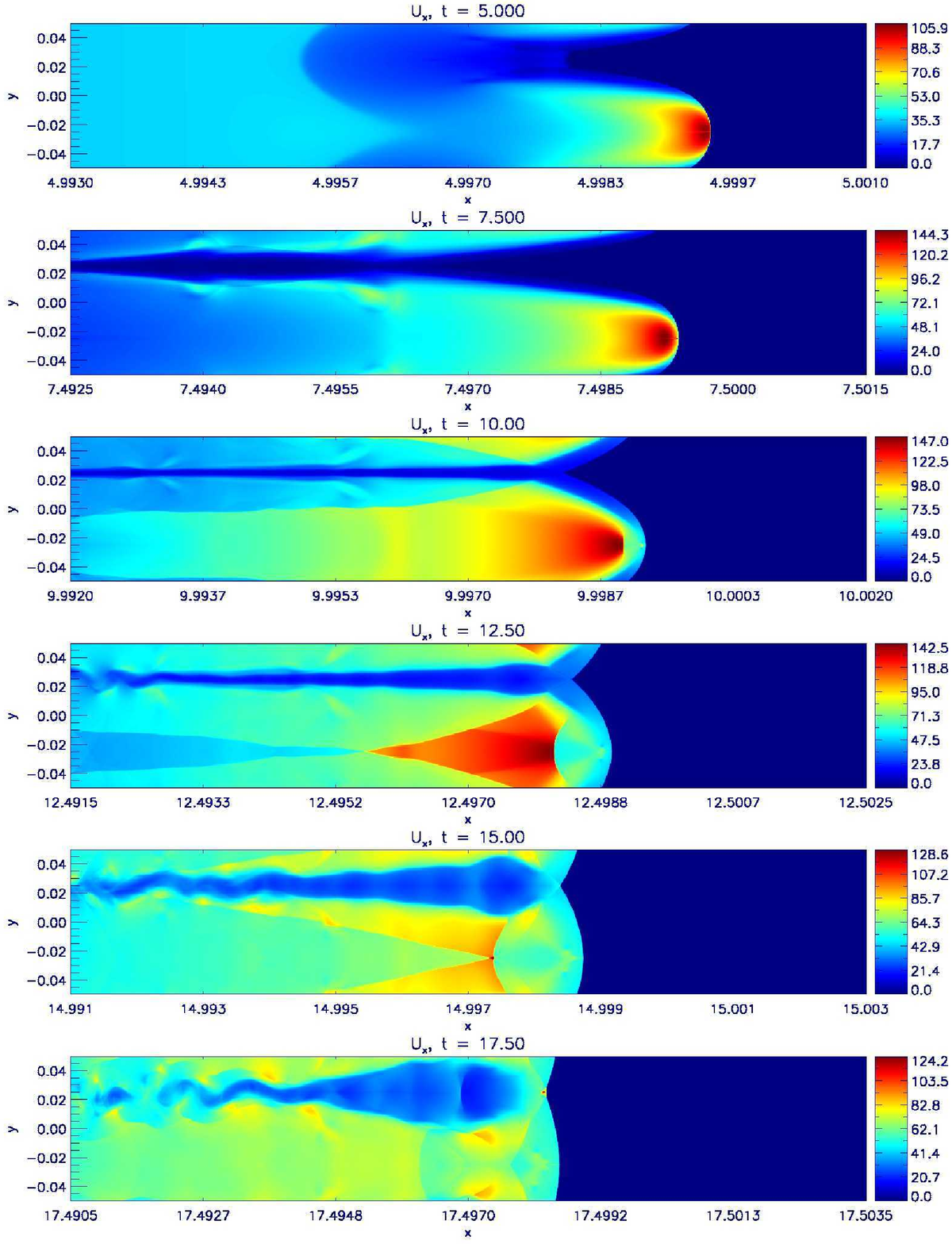}\caption{\footnotesize The parallel 4-velocity $u_x$ for $k\approx4.8\cdot10^2k_0$ and $\varepsilon\approx2$.}\label{r2e1BIS_Ux_01-02-03-04-05-06-07-08-09-10-11_a}
\end{center}\end{figure}
\begin{figure}[!ht]\begin{center}
\epsscale{.8}\plotone{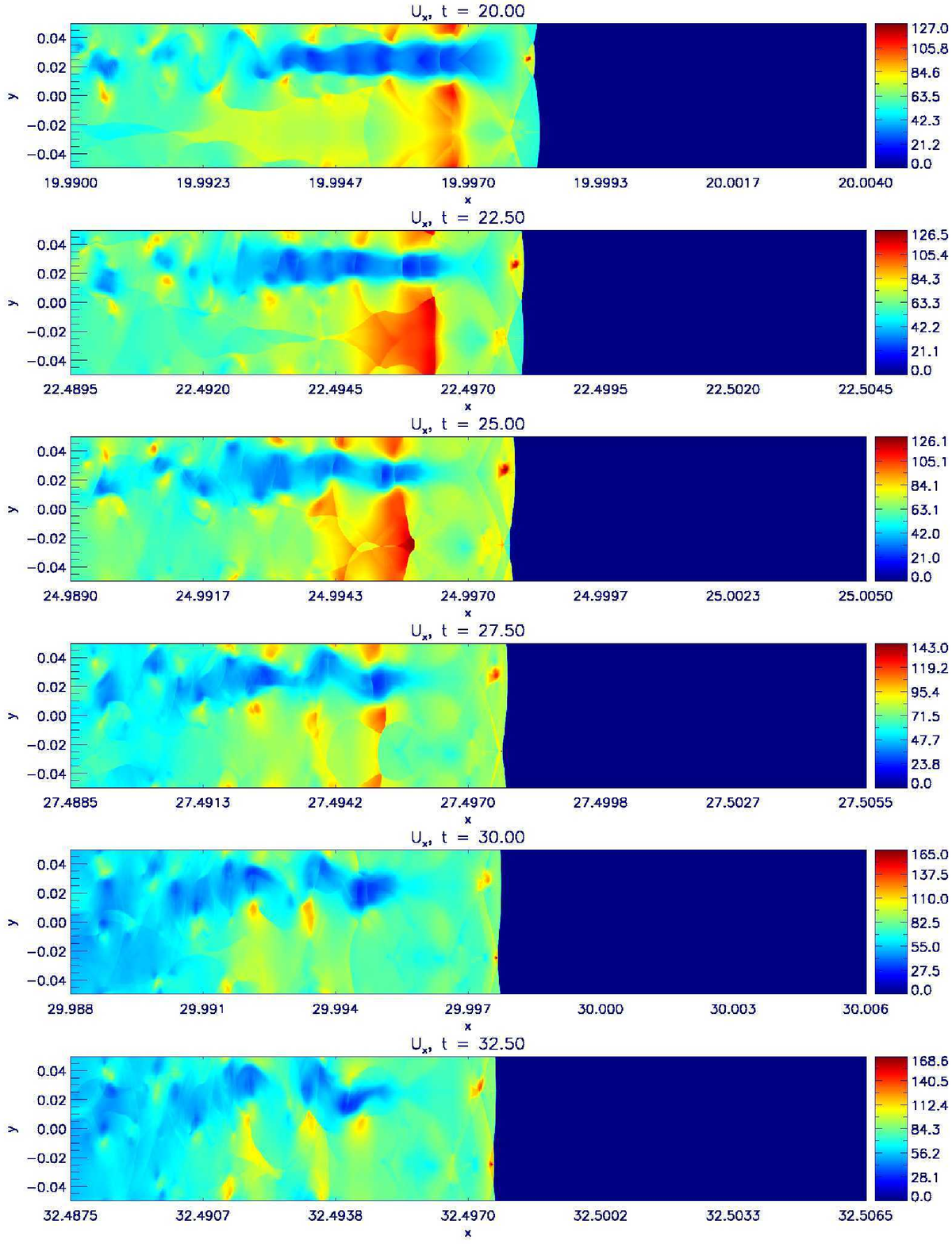}\caption{\footnotesize The parallel 4-velocity $u_x$ for $k\approx4.8\cdot10^2k_0$ and $\varepsilon\approx2$.}\label{r2e1BIS_Ux_01-02-03-04-05-06-07-08-09-10-11_b}
\end{center}\end{figure}
\clearpage

\begin{figure}[!ht]\begin{center}
\epsscale{.8}\plotone{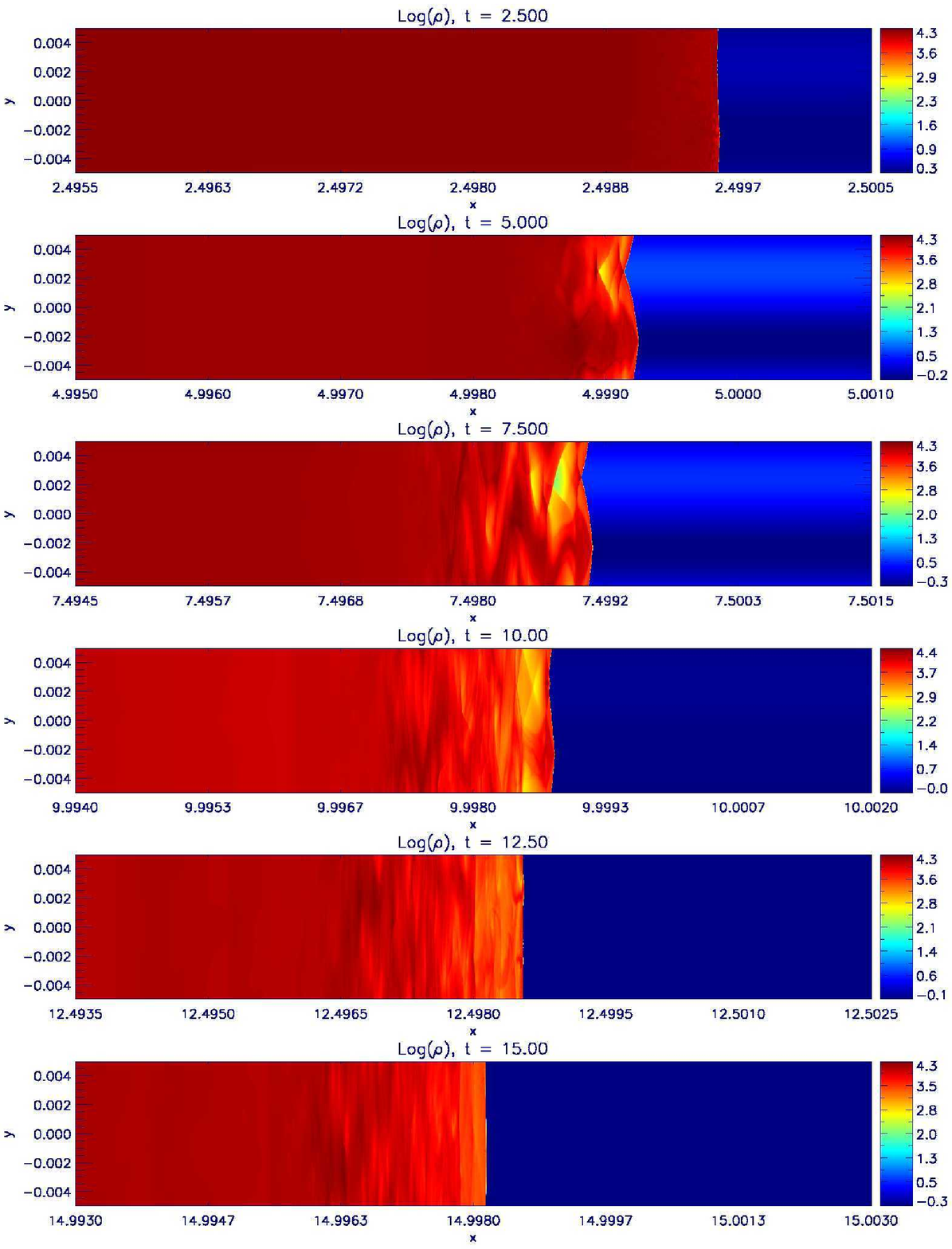}\caption{\footnotesize The density logarithm for $k\approx4.8\cdot10^3k_0$ and $\varepsilon\approx0.5$.}\label{r2e2_rho_00-01-02-03-04-05}
\end{center}\end{figure}

\begin{figure}[!ht]\begin{center}
\epsscale{.8}\plotone{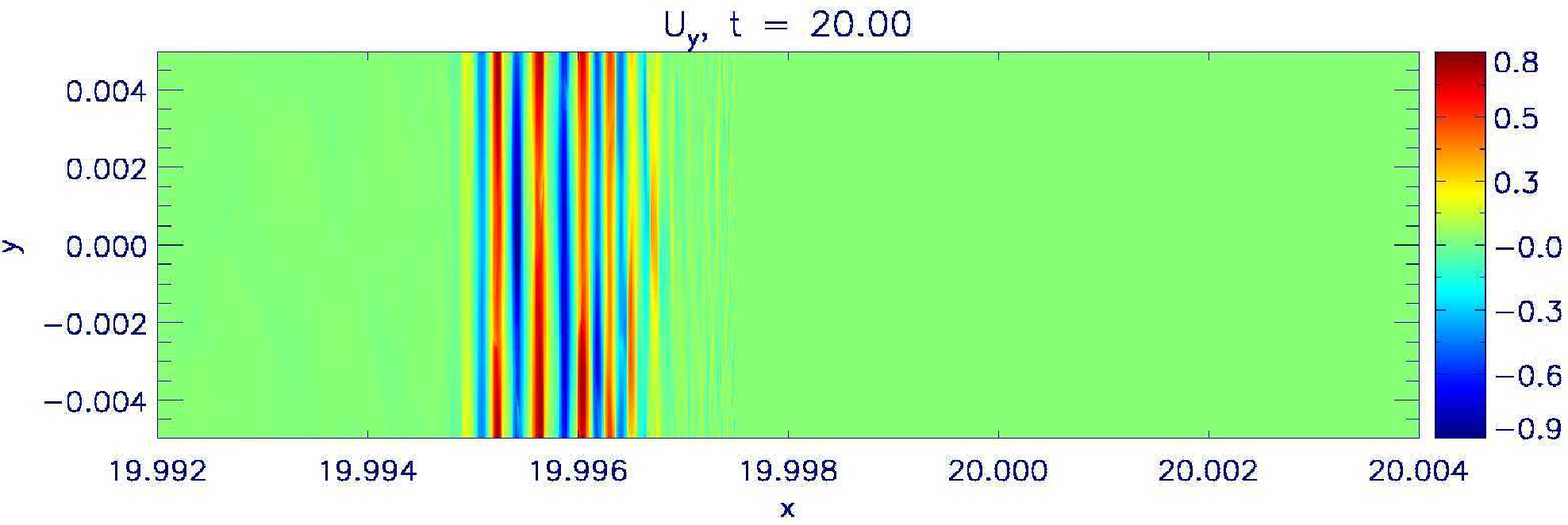}\caption{\footnotesize The transverse 4-velocity $u_y$ for $k\approx4.8\cdot10^3k_0$ and $\varepsilon\approx0.5$.}\label{r2e2_uy_07}
\end{center}\end{figure}

We also include some snapshots of two other simulations showing close analogies with the previous one. The first and most spectacular one deals with $k\approx4.8\cdot10^2k_0$ and $\varepsilon\approx2$, and presents clear evidences of Kelvin-Helmotz instabilities; compare Figs.~\ref{r2e1BIS_rho_01-02-03-04-05-06-07-08-09-10-11_a} -- \ref{r2e1BIS_rho_01-02-03-04-05-06-07-08-09-10-11_b} and \ref{r2e1BIS_Ux_01-02-03-04-05-06-07-08-09-10-11_a} -- \ref{r2e1BIS_Ux_01-02-03-04-05-06-07-08-09-10-11_b}: at early times, behind the valley, due to the stopping presence of the overdensity, a thin, slow layer courses the unperturbed, fast flow, giving rise to the instability. The second one, realized with $k\approx4.8\cdot10^3k_0$ and $\varepsilon\approx0.5$, quickly comes to a restoring of the shock surface planarity (Fig.~\ref{r2e2_rho_00-01-02-03-04-05}). 
    The complexity of the fluctuation pattern described above can be seen from this last simulation: Fig.~\ref{r2e2_uy_07} shows the tightly arranged warp of $u_y$ already at an intermediate evolutionary phase.

\clearpage
\clearpage

\subsection{Cylindrical clouds}{}\label{sfec}
The perturbations we are going to deal with here complement the ones we explored above. Indeed, if the sinusoidal bars provide a quite thorough description of those phenomena occurring during the sweeping up of a smoothly inhomogeneous upstream, a cylindrical overdensity can well represent the sharp contrast of a typical cloud in a clumpy circumburst medium.

We present here the results of a simulation with a homogeneous cylindrical cloud (with axis parallel to the $z$-axis) of radius $r\approx2.4\cdot10^{-1}k_0^{-1}$ placed in the upstream of the usual planar shock. The cloud density is larger by a factor $10^3$ than the unperturbed upstream one.

In order to save computational time, we considered only half cloud and substituted periodic $y$-boundary conditions with reflective (rigid walls) ones.

\begin{figure}[!ht]\begin{center}
\epsscale{.8}\plotone{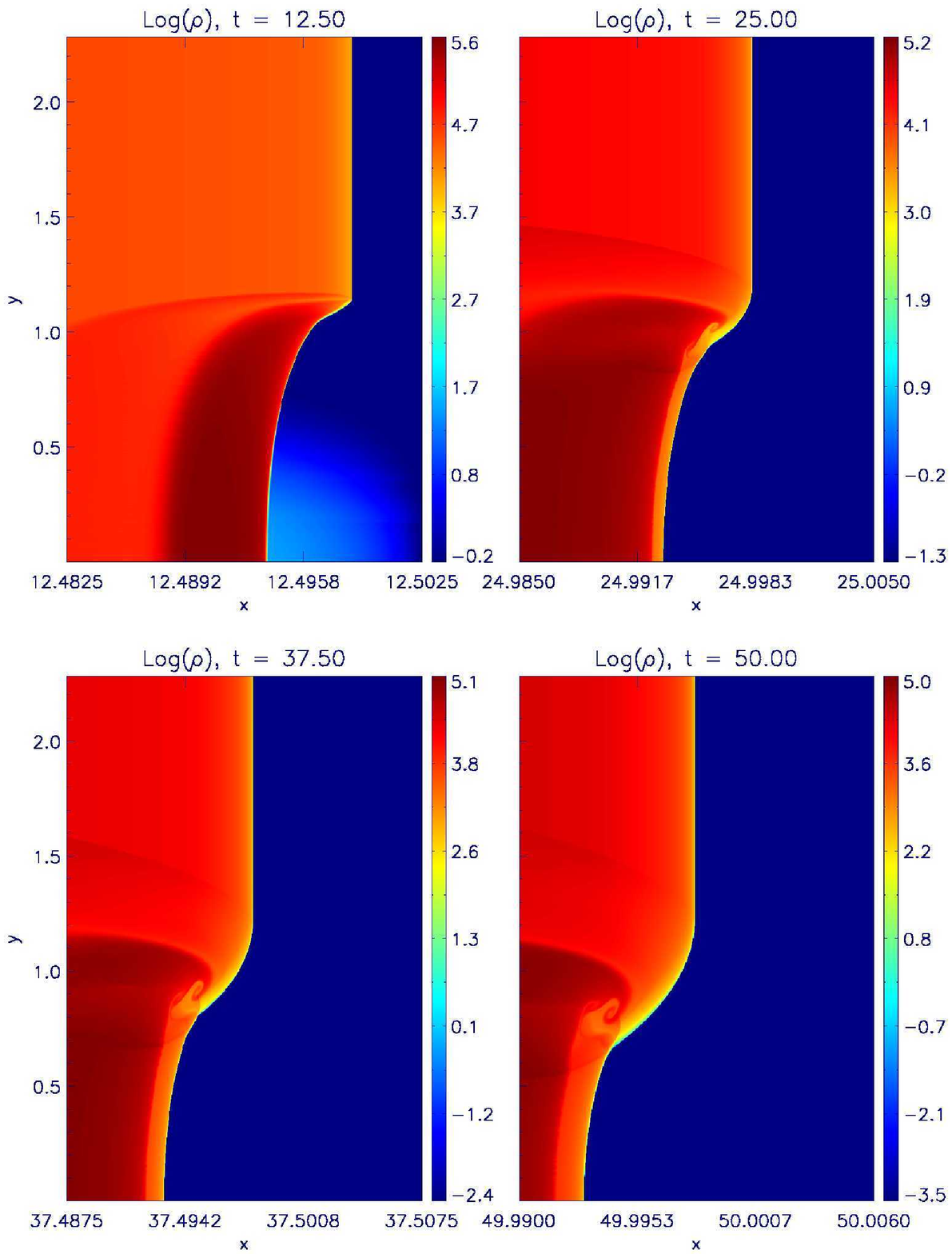}\caption{\footnotesize The density logarithm for $r\approx2.4\cdot10^{-1}k_0^{-1}$ and $\varepsilon\approx3$.}\label{r5a3A4_rho_04-09-14-19}
\end{center}\end{figure}

\begin{figure}[!ht]\begin{center}
\epsscale{.8}\plotone{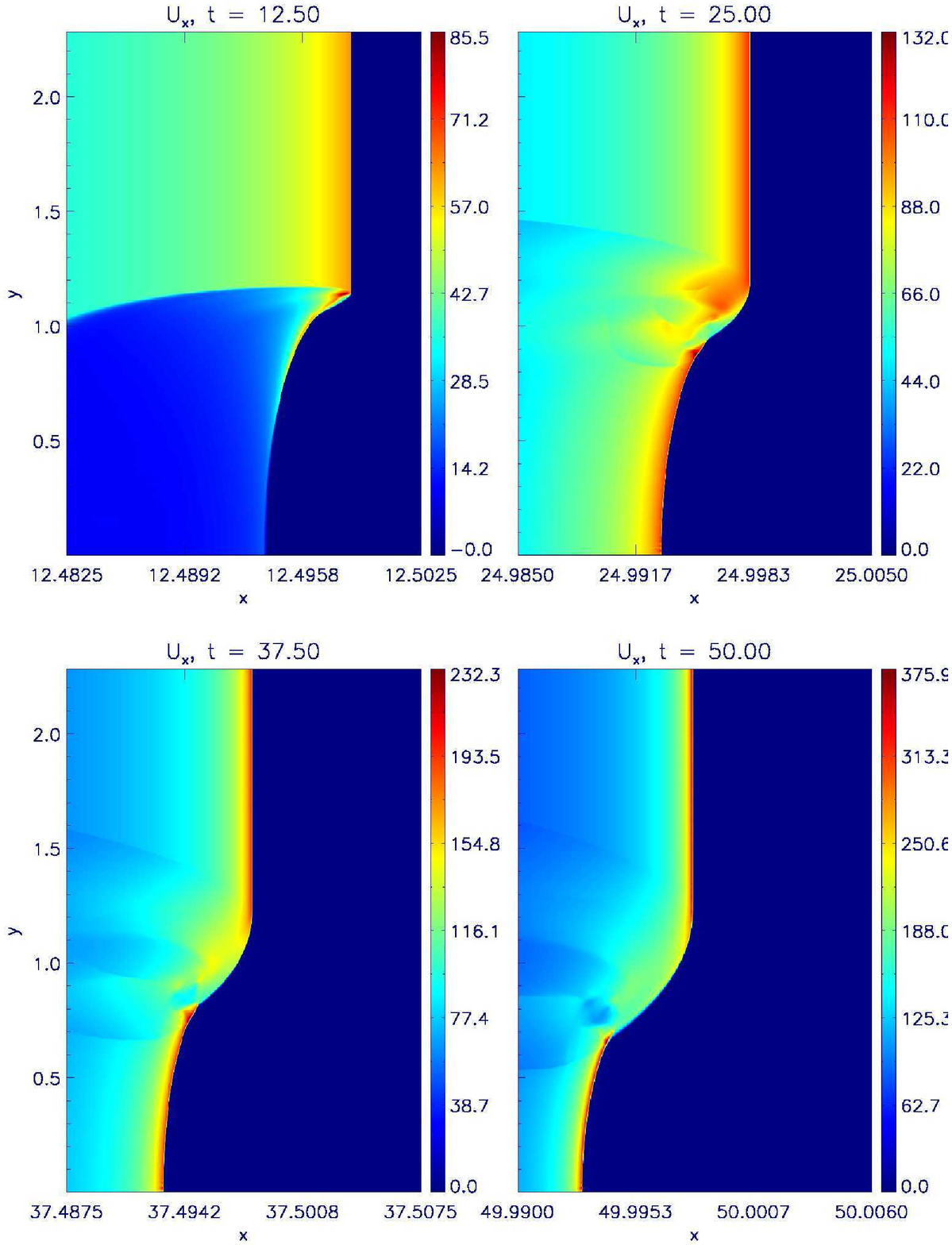}\caption{\footnotesize The parallel 4-velocity $u_x$ for $r\approx2.4\cdot10^{-1}k_0^{-1}$ and $\varepsilon\approx3$.}\label{r5a3A4_ux_04-09-14-19}
\end{center}\end{figure}

\begin{figure}[!ht]\begin{center}
\epsscale{.8}\plotone{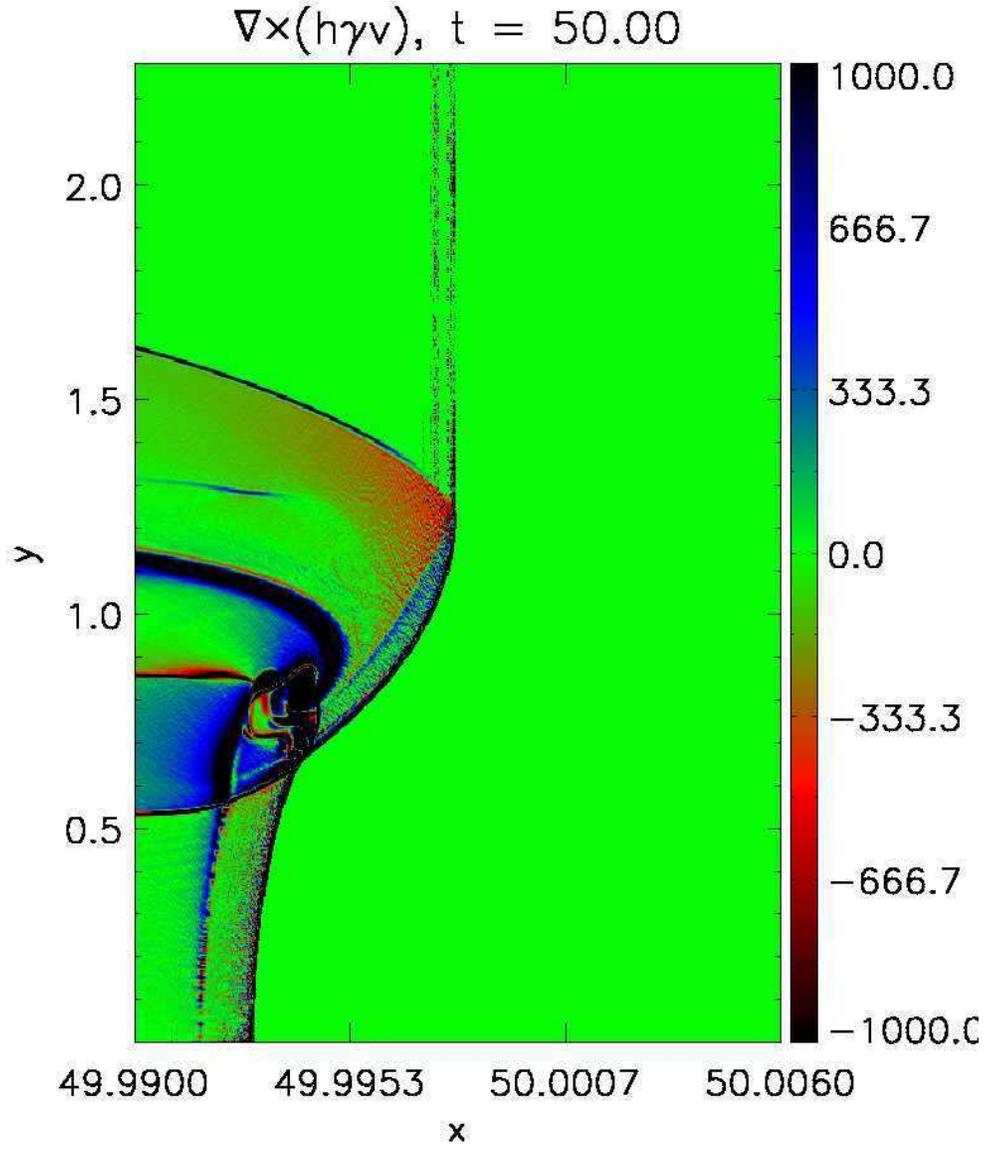}\caption{\footnotesize The vorticity $\vec{\nabla}\times(h\gamma\vec{v})$ for $r\approx2.4\cdot10^{-1}k_0^{-1}$ and $\varepsilon\approx3$.}\label{r5a3A4_vort_19}
\end{center}\end{figure}

Figs.~\ref{r5a3A4_rho_04-09-14-19} and~\ref{r5a3A4_ux_04-09-14-19} show the main evolution phases of the system: closely related to the first widest sinusoidal case, the unperturbed shock tends to fill the valley from the boundary of the cloud. However, in this case, because of the steeper wrinkle in the shock, the non-adiabaticity of the flow across the discontinuity surface allows a vortex to develop and to be advected downstream (Fig.~\ref{r5a3A4_vort_19}). The study of the vortex dynamics and its relevance with regard to GRBs physics will be discussed in a forthcoming paper.

\clearpage
\clearpage

\section{Turbulent ambient density}{}\label{par5}
We discussed before the evolution of a shock in a self-similar regime under the effect of some perturbing agent. 
We now wish to investigate how fast such a self-similar regime is reached.

\begin{figure}[!ht]\begin{center}
\epsscale{.8}\plotone{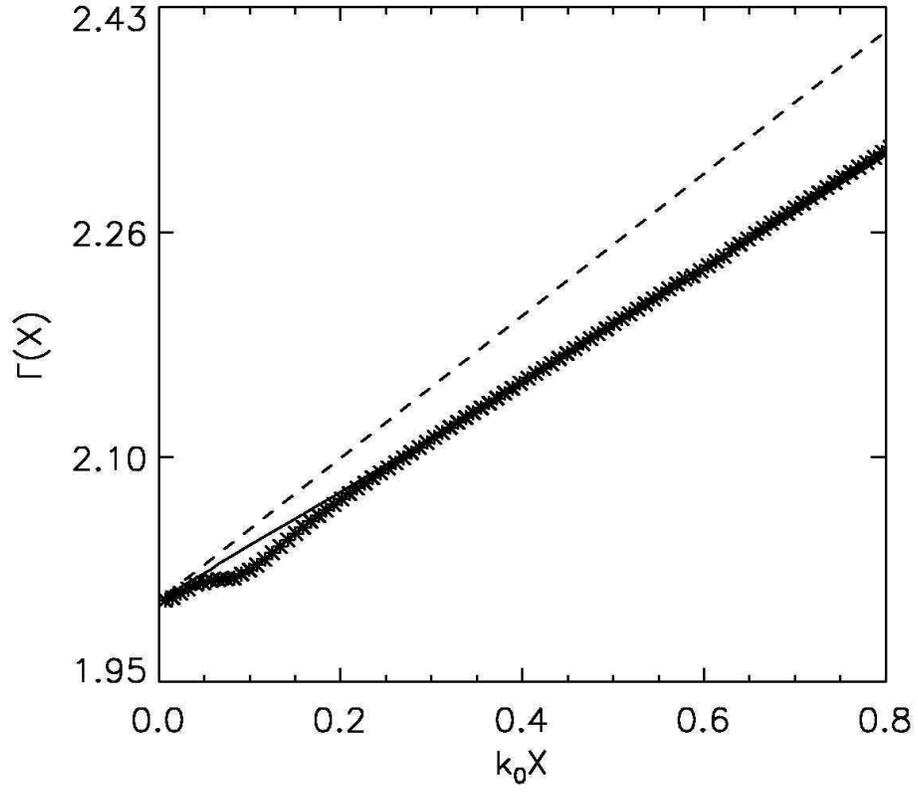}\caption{\footnotesize Shock Lorentz factor evolution as function of the distance traveled into the exponential atmosphere for $\Gamma_0=2$ (stars for simulation, solid line for self-similar prediction, dashed line for the hyperrelativistic limit of the latter).}\label{auto2}
\end{center}\end{figure}

\begin{figure}[!ht]\begin{center}
\epsscale{.8}\plotone{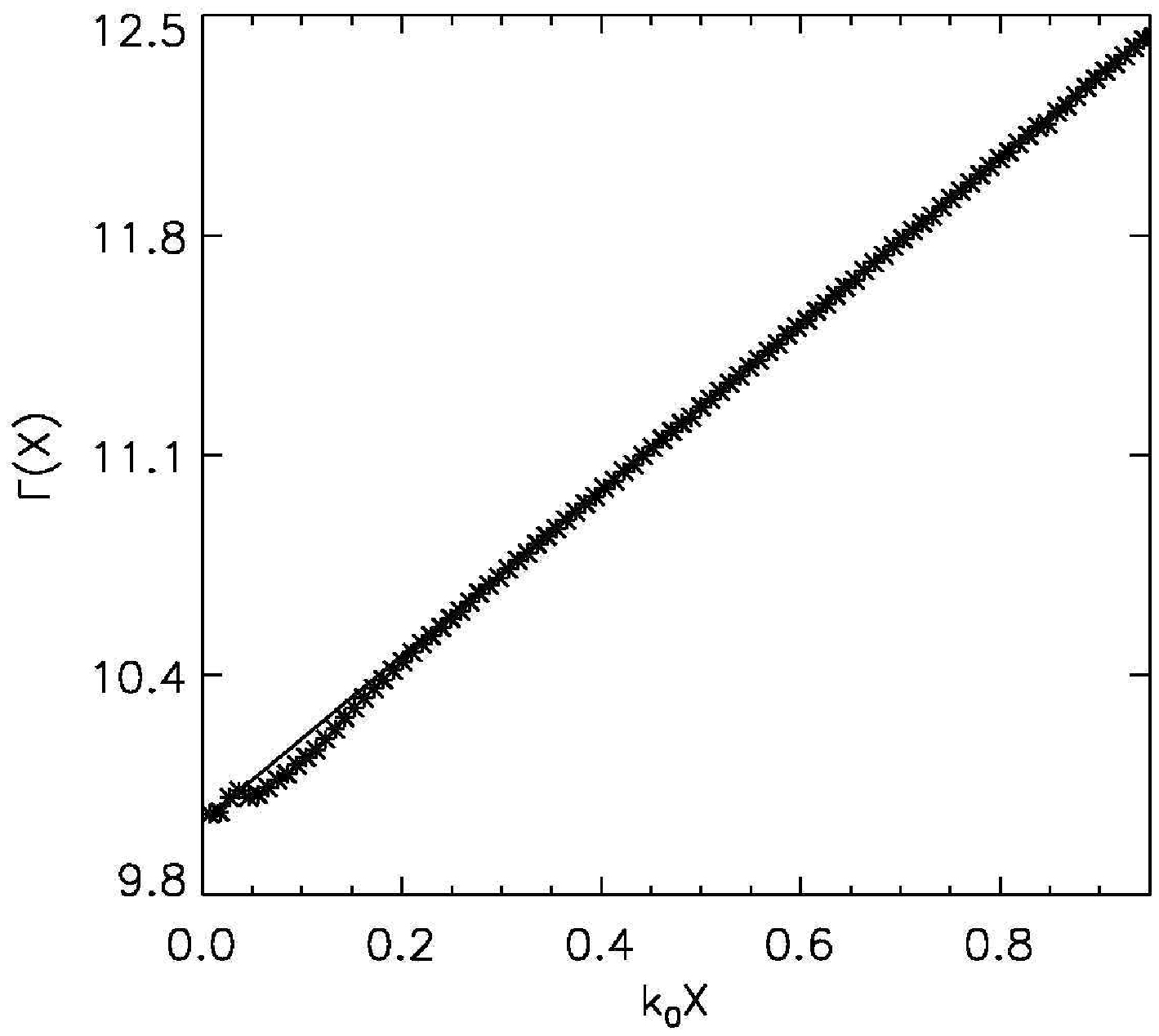}\caption{\footnotesize Shock Lorentz factor evolution as function of the distance traveled into the exponential atmosphere for $\Gamma_0=10$ (stars for simulation and solid line for self-similar prediction).}\label{auto10}
\end{center}\end{figure}

\begin{figure}[!ht]\begin{center}
\epsscale{.8}\plotone{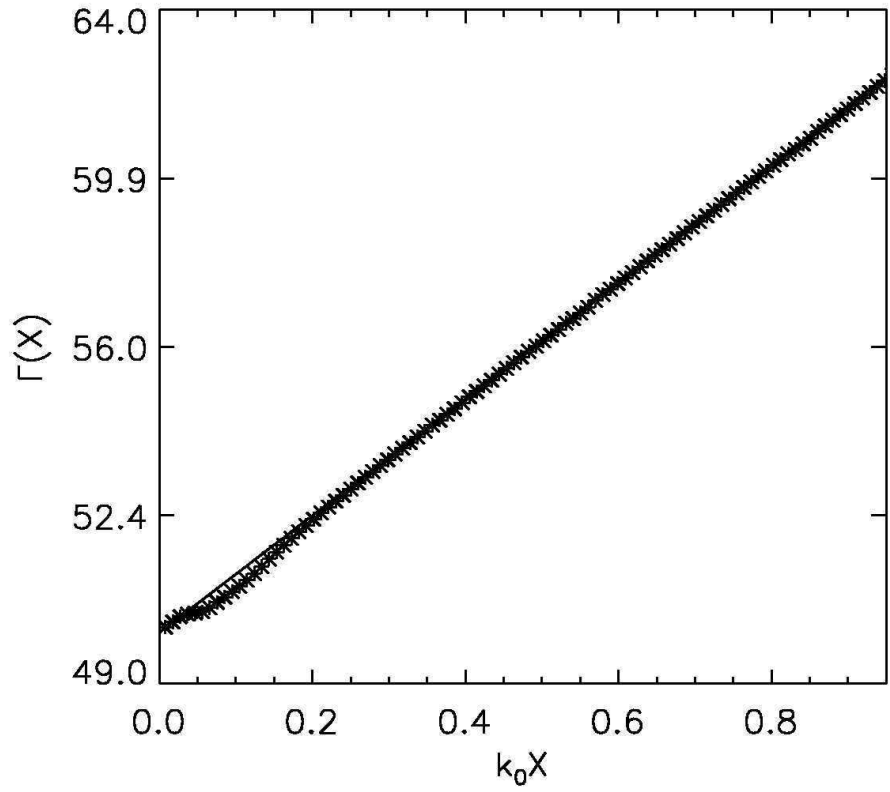}\caption{\footnotesize Shock Lorentz factor evolution as function of the distance traveled into the exponential atmosphere for $\Gamma_0=50$ (stars for simulation and solid line for self-similar prediction).}\label{auto50}
\end{center}\end{figure}

Let us consider the usual shock with initial Lorentz factor $\Gamma_0$ which, at $t=0$, comes out of a homogeneous upstream and starts to propagate into an exponential atmosphere. In Figs.~\ref{auto2} -- \ref{auto50} we compare (for three different values of $\Gamma_0$ spanning a wide range of relativistic regimes) the simulated Lorentz factor as a function of the length scale fraction covered by the shock with the prediction given by the self-similar theory (\rf{shloc}).
\clearpage
\clearpage

It is quite a significant fact that self-similarity is reached almost immediately after the exponential length scale switch (in our scheme, the length scale of the atmosphere switches from $\infty$ for $x<0$ to $k_0^{-1}$ for $x>0$), with a very weak dependence on $\Gamma_0$: one senses that larger Lorentz factors help the shock to follow closer the self-similar run of the acceleration in the first 15 -- 20\% of length scale after the atmosphere transition.

We now show that such a property allows us to consider the shock behavior as uniquely ruled by atmosphere density value at the point of interest. Obviously the II type nature of this self-similar problem will play a fundamental role here, since we are actually requiring the shock to have an infinite piston behind it which compensates for the indefinite energy supply in the upstream. 

Let us consider a generic upstream atmosphere density profile $\Pi(x)$. We assume $\Pi(x)$ to be a monotonic decreasing function. It is possible to approximate $\Pi(x)$ with a finely broken line made up by a set of segments of exponentials. At least as self-similarity is reached on temporal -- and thus spatial -- scales smaller than the spatial scales required by $\Pi(x)$ to appreciably depart from the local tangent exponential, one is allowed to determine the infinitesimal increases of $\Gamma$ by means of \rf{shloc}: \begin{equation}\Gamma+d\Gamma=\Gamma\left(\frac{\Pi+d\Pi}{\Pi}\right)^{1/\alpha}\;.\label{sbs}\end{equation} Integrating step by step \rf{sbs}, the generalization to an arbitrary density profile of \rf{shloc} is easily obtained: \begin{equation}\Gamma\approx\Gamma_i\left(\frac{\Pi}{\Pi_i}\right)^{1/\alpha}\,.\label{slg}\end{equation} As a result, shock speed will depend only on the initial Lorentz factor and on the local density value.

Having this fact on our mind, we can calculate how much a shock propagating in an atmosphere $\Pi(x)$ will fall behind an initially identical one (also $\Pi_0=\rho_0$) which propagates -- unperturbed -- in the usual exponential profile $\rho(x)$. Calling $X_1$ the former's position and $X_0$ the latter's one and imposing $X_0(t=0)=X_1(t=0)=0$: \begin{equation}\dot{X_0}=v\left(\Gamma_0\exp-\frac{k_0X_0}{\alpha}\right)\approx1-\frac{\exp\frac{2k_0t}{\alpha}}{2\Gamma_0^2}\;;\label{ics0}\end{equation}\begin{equation}\dot{X_1}=v\left(\Gamma_0\left(\frac{\Pi(X_1)}{\rho_0}\right)^{1/\alpha}\right)\approx1-\frac{\left(\frac{\rho_0}{\Pi(t)}\right)^{2/\alpha}}{2\Gamma_0^2}\;.\label{ics1}\end{equation} Subtracting \rf{ics0} from \rf{ics1}, one obtains the equation for the delay: \begin{equation}\frac{d}{dt}\Delta X\approx\frac{1}{2\Gamma_0^2}\left[\exp\frac{2k_0t}{\alpha}-\left(\frac{\rho_0}{\Pi(t)}\right)^{2/\alpha}\right]\;.\label{milst}\end{equation}

\rf{milst} is an interesting result, since it may prove to be useful for accurately estimating the shock position in an arbitrary atmosphere without wasting any time in full blown simulations.

As an example, let us write the density profile as \begin{equation}\Pi(x)=\rho(x)\left(1+\varepsilon\delta\Pi(x)\right)\;.\end{equation} If we are dealing with a slightly perturbed upstream, $\varepsilon$ will be much smaller than 1, thus justifying the following simplifications to \rf{milst}: \begin{equation}\label{milst2}\frac{d}{dt}\Delta X\approx\frac{1}{2\Gamma_0}\exp\frac{2k_0t}{\alpha}\left\{1-\left[1+\varepsilon\delta\Pi(t)\right]^{-\frac{2}{\alpha}}\right\}\approx\frac{\varepsilon\delta\Pi(t)\exp\frac{2k_0t}{\alpha}}{\alpha\Gamma_0^2}\;.\end{equation} Let us suppose now that a planar shock encounters a turbulent upstream. According to the approximation of independent evolution of each flow cylinder we extensively discussed above, we can derive the statistical properties of the shock wrinkles at each $x$ (or, equivalently, at each $t$). As a starting point we can imagine that each upstream cylinder perturbation (identified by a pair $(y,z)$) is a particular realization of a power spectrum $A(k)$, such that $A(k)=\left|\widetilde{\delta\Pi}(k,y,z)\right|,\forall y,z$: \begin{equation}\delta\Pi(x,y,z)=\int_0^{\infty}dkA(k)\cos\left(kx+\delta(k)\right)\;,\label{ft}\end{equation} where the phase $\delta(k)$ is a random variable determining each realization with a probability distribution given by \begin{equation} P\left(\delta(k)\right)=\frac{H\left(\delta(k)\right)H\left(2\pi-\delta(k)\right)}{2\pi}\;.\end{equation} Integrating \rf{milst2} \begin{equation}\Delta X(t)=\int_0^t\frac{\varepsilon\delta\Pi(\tau)\exp\frac{2k_0\tau}{\alpha}d\tau}{\alpha\Gamma_0^2}\end{equation} and substituting \rf{ft} we obtain \begin{equation}\Delta X(t)=\int_{\kappa}^K\frac{\varepsilon A(k)}{\alpha\Gamma_0^2}\int_0^t\cos\left(k\tau+\delta(k)\right)\exp\frac{2k_0\tau}{\alpha}d\tau dk\;.\label{somma}\end{equation} Here two cuts have been introduced in order to exclude from the computation turbulence wave-length larger than $x\approx t$ itself (infrared cut $\kappa\equiv t^{-1}$) or smaller than the scales reached by transverse diffusive phenomena smoothing out high wave-number wrinkles (ultraviolet cut $K\equiv25\Gamma t^{-1}$ -- see \rf{tsm}). From \rf{somma} it is obvious that $\Delta X$ is itself a random variable given by the sum of (infinite) random variables, each of them identified by the parameter $k$ and depending on the random phase $\delta(k)$. The central limit theorem completely characterizes (from a statistical point of view) $\Delta X$ as a random variable Gaussian distributed with an average over the ensemble of cylinders $\langle\Delta X(t)\rangle=0$ and a variance $\sigma\equiv\langle\Delta X^2\rangle$ given by the (infinite) sum of the variances $d\sigma(k)$: \begin{equation}\frac{d\sigma}{dk}\equiv\frac{\int_0^{2\pi}\left[\frac{\varepsilon A(k)}{\alpha\Gamma_0^2}\int_0^t\cos(k\tau+\delta)\exp\frac{2k_0\tau}{\alpha}d\tau\right]^2d\delta}{2\pi}\;,\end{equation} whence, summing over $k$, it results: \begin{equation}\label{char}\langle\Delta X^2\rangle(t)=\varepsilon^2\int_{\kappa}^K\frac{A^2(k)\left[1+\exp\frac{4k_0t}{\alpha}-2\cos(kt)\exp\frac{2k_0t}{\alpha}\right]}{2\Gamma_0^4(\alpha^2k^2+4k_0^2)}dk\;.\end{equation}

The characterization of the shock position distribution provided by \rf{char} may represent a benchmark for several applications. Minor modifications should be sufficient, for instance, to obtain an estimate of the features in the fluctuations in the afterglow light curve, provided a model of upstream turbulence is given. Such a kind of study is considered of great interest, particularly in light of the recent \emph{Swift} observations: bumps, flares and plateaus have often been observed in place of smooth
power-law decays, thus challenging our understanding of the afterglow production.
\section{Conclusions}{}\label{concl}
We started by testing whether the PLUTO code is appropriate for treating the evolution of hyperrelativistic shock waves with Lorentz factors even in excess of $10^2$, obtaining satisfactory evidence of coherency with self-similar theory developed by~\citet{perna} and~\citet{palma} for shock acceleration in exponential atmosphere.

To follow we tried to answer the question on how nonlinear effects may let the instability evolution depart from the linear behavior. We studied several perturbing agent configurations, concluding that the shock will tend to restore the original planar shape of the discontinuity surface on a time scale given by $T_{sm}\sim25\Gamma k^{-1}$. We intend to remark that such a behavior does not appear as a typical saturation phenomenon, due to the lack of a competition between a destabilizing factor (which is actually missing) and a restoring agent. The reason is easily found: while in the Newtonian counterpart of the problem the destabilizing factor is given by the tendency of the zero-th order solution to preserve any speed difference between adjacent flow columns (even better, it grows indefinitely; see~\cite{chev1}), here the acceleration is in terms of an homogeneous growth of the Lorentz $\Gamma$ factor. As a consequence, in the hyperrelativistic regime, essentially due to the existence of the speed limit $c$, even in absence of restoring effects, the maximum gap that a difference of $\Gamma$ between two distinct cylinders can produce is \begin{equation}\Delta X_{\max}\approx\frac{\alpha(\Gamma_2^2-\Gamma_1^2)}{4k_0\Gamma_1^2\Gamma_2^2}=k_0^{-1}O(\Gamma^{-3})\;.\end{equation} This explains the lack of a substantial destabilizing factor: even if two regions of the shock -- for example as a consequence of an inhomogeneous upstream -- travel with different speeds and are at different positions, as soon as the perturbing agent disappears, they will tend to reach sooner or later a maximum gap. At that point the only acting process is the smoothening influence of the secondary shocks. As a result, the shock will tend inexorably to the zero-th order solution (in other terms, the saturation point is that of no perturbation!).

In the last section we found that self-similarity is reached almost immediately in the flow and concluded that this allows us to predict the shock position as a function of the only initial and final upstream density. We applied such a result to the case of a turbulent ambient density and derived an analytical expression for the dispersion of the shock positions at different transverse positions.

The subject we treated in this paper is expected to have a great relevance with regard to models describing GRB radiation, especially those concerning the afterglow emission at the external forward shock. The recent \emph{Swift} observations have shown lots of bump, flares and plateaus in place of smooth power-law decays in the light-curves~\citep{meszaros}, thus challenging our understanding of the afterglow production. \citet{lazzati,heyl,nakar} suggested that such a variability in the light-curves may be the result of the presence of density bumps in the upstream medium. The upcoming launch of GLAST is likely to provide an even more detailed description of such events, thus requiring a more accurate modeling of the underlying physics. It goes without saying, therefore, that having a good theory for the dynamics of highly relativistic shock waves represents a key point of our capability to properly predict the emission expected from an afterglow. 

Future developments of this work will involve the evolution of shocks propagating in a magnetized ambient medium. An easy and straightforward extension of the numerical setup hitherto developed would dispel the uncertainties greatly affecting at the moment this theoretical issue and may provide unambiguous evidences on a hot topic of the GRB theory.

\end{document}